\documentclass[a4paper,11pt]{article}
\usepackage[utf8]{inputenc}

\usepackage{bm}
\usepackage{comment} 

\usepackage[top=15truemm,bottom=20truemm,left=20truemm,right=20truemm]{geometry}
\usepackage[colorlinks=true,urlcolor=blue,anchorcolor=black,citecolor=blue,linkcolor=black,filecolor=black,menucolor=black,pagecolor=black,linktocpage=true,pdfproducer=medialab,pdfa=true]{hyperref}
\usepackage{graphicx}
\usepackage{amsmath,latexsym,amssymb,mathrsfs,ascmac,physics,mathtools}
\numberwithin{equation}{section}
\usepackage[affil-it]{authblk}
\usepackage{cite}

\title{Wormholes as perturbations of near-horizon black hole geometries: no-go theorems within effective field theories}

\author[1]{Takamasa Kanai \thanks{kanai@kochi-ct.ac.jp}}
\affil[1]{Department of Social Design Engineering,
National Institute of Technology (KOSEN), Kochi College,
200-1 Monobe Otsu, Nankoku, Kochi, 783-8508, Japan}

\author[2]{Kengo Maeda \thanks{maeda302@sic.shibaura-it.ac.jp}}
\affil[2]{Faculty of Engineering, Shibaura Institute of Technology, Saitama 330-8570, Japan}

\author[3]{Daisuke Yoshida \thanks{dyoshida@math.nagoya-u.ac.jp}}
\affil[3]{Department of Mathematics, Nagoya University, Nagoya 464-8602, Japan}

\date{}

\begin{document}

\maketitle 

\begin{abstract}
We reformulate the construction of wormhole solutions as perturbations around near-horizon geometries of near-extremal Reissner--Nordstr\"{o}m  black holes in four dimensions and equal-angular-momenta Myers--Perry black holes in five dimensions.  When the negative Casimir energy is taken as the source, this framework reduces to the Maldacena--Milekhin--Popov construction for magnetically charged wormholes. We then show that, in contrast, such perturbative constructions cannot be realized within the effective field theory approach to higher-derivative corrections. Remarkably, this conclusion holds irrespective of the specific form of the correction terms. The key observation is that the enhanced symmetries in the near-horizon region severely constrain the effective energy--momentum tensor near the throat. This prevents the formation of the traversable throat structure. Our analysis therefore establishes no-go theorems: traversable wormholes cannot arise perturbatively from Reissner--Nordstr\"{o}m or Myers--Perry black holes in an effective field theory approach. Their realization would require either new ingredients, such as Casimir energy, or black holes with reduced symmetry.
\end{abstract}

\tableofcontents

\newpage 
\section{Introduction}
A central question in gravitational physics beyond the purely classical regime is what principles determine the limits of possible spacetime geometries. This issue becomes particularly relevant once the null convergence condition, which is equivalent to the null energy condition through the Einstein equations, is no longer assumed.
Although this condition is generally expected to capture essential properties of General Relativity with classical fields, quantum effects can lead to its violation. This raises the broader issue of how the boundaries between physical and unphysical spacetimes should be defined in semiclassical or quantum gravity.

Alternatives to the null energy condition in the semiclassical regime have been proposed and discussed in the literature. One direction involves considering averaged energy conditions, leading to the (achronal) averaged null energy condition \cite{PhysRevD.17.2521, Borde:1987qr,Wald:1991xn,Graham:2007va} or the quantum energy inequality \cite{Ford:1994bj, Ford:1990id,Fewster:1998pu,Kontou:2020bta}.
Another approach is motivated by the extension of the thermodynamic properties and entropy-based arguments of classical black holes \cite{Bekenstein:1973ur, Bardeen:1973gs, Hawking:1975vcx, Wald:1999vt} to the semiclassical regime. For example, the quantum focusing conjecture \cite{Bousso:2015mna} generalizes the role of the null convergence condition by imposing constraints on the evolution of generalized entropy along null congruences. 

A systematic framework to study the properties of spacetimes incorporating effects from microscopic physics is effective field theory (EFT)~\cite{Weinberg:1978kz, Georgi:1993mps,Donoghue:1994dn,Donoghue:1995cz,Burgess:2003jk}, which takes into account all symmetry-allowed terms in the action, organized in an expansion based on the mass dimension of the operators. Unlike modified gravity theories, such as $f(R)$ gravity~\cite{Nojiri:2006ri,Sotiriou:2008rp}, which often redefine the gravitational dynamics nonperturbatively, EFT treats higher-derivative corrections perturbatively on top of Einstein gravity, providing a controlled expansion valid below a certain cutoff scale. Such corrections can violate the null convergence condition, making them natural candidates to probe the limits of spacetime structures beyond classical general relativity. In addition, consistency requirements, such as positivity bounds \cite{Adams:2006sv, Cheung:2014ega, Cheung:2016wjt, Hamada:2018dde, Tokuda:2020mlf}, arising from ultraviolet physics further restrict the space of allowed EFT corrections, pointing toward deeper principles governing spacetime physics. In our earlier work~\cite{Kanai:2024zsw}, we estimated the cutoff scale of the gravitational EFT required for the validity of the quantum focusing conjecture (see also Refs.~\cite{Fu:2017lps,Leichenauer:2017bmc}).
These considerations highlight the role of the EFT approach as a crucial framework linking consistency conditions from ultraviolet physics with the possible realizations of spacetime geometries.

Wormholes provide a particularly sharp test case for these considerations, since their existence is tightly constrained by energy conditions. This has been illustrated by explicit solutions supported by phantom scalar fields, such as the Bronnikov--Ellis wormhole \cite{Bronnikov:1973fh, Ellis:1973yv} and its subsequent developments such as asymmetric generalization \cite{Armendariz-Picon:2002gjc} and multi-leg configurations \cite{Makita:2025bao}. On the one hand, Morris and Thorne analyzed static, spherically symmetric wormhole geometries directly and showed that the flaring-out condition at the throat necessarily requires violations of the null convergence condition~\cite{Morris:1988cz}. The relation between the existence of wormholes and the violation of a kind of energy or convergence conditions has been developed into a general theorem: the topological censorship theorem \cite{Friedman:1993ty, Friedman:2006tea} establishes that no ``short'' traversable wormholes can exist under the achronal averaged null energy condition \cite{Graham:2007va}. 

Beyond general relativity with classical matter, however, numerous attempts have been made to realize wormhole geometries in frameworks where the null convergence conditions can be relaxed. Modified gravity theories, such as higher-derivative extensions of Einstein gravity or models including new dynamical ingredients coupled to gravity in a nontrivial way, have been explored as candidate frameworks in which the null convergence condition may be violated.
For example, wormhole solutions have been studied in $f(R)$ gravity~\cite{Lobo:2009ip, Harko:2013yb,Mishra:2021ato,Parsaei:2021wrw,Sokoliuk:2022xcf}, dilaton--Gauss--Bonnet gravity~\cite{Kanti:2011jz, Kanti:2011yv}, scalar-tensor theory such as Horndeski and beyond Horndeski theory~\cite{Evseev:2017jek,Franciolini:2018aad,Mironov:2018pjk,Mironov:2018uou,Bakopoulos:2021liw}
, and higher dimensional Gauss--Bonnet/ Lovelock gravity~\cite{Bhawal:1992sz, Maeda:2008nz, Mehdizadeh:2015jra, Mehdizadeh:2015dta, Giribet:2019dmg, Mehdizadeh:2021kgv}. In addition, even within the classical general relativity, wormhole solutions supported by the energy of a fermionic field have been discussed \cite{Blazquez-Salcedo:2020czn, Bolokhov:2021fil, Konoplya:2021hsm, Danielson:2021aor}. Also, wormhole connecting asymptotic AdS regions has been studied \cite{Anabalon:2018rzq}.

Complementary to these approaches, traversable wormholes supported by the energy of quantum field has been studied so far. For example, traversable wormholes in holographic setups have been extensively studied in Refs.~\cite{Gao:2016bin,Maldacena:2017axo,Maldacena:2018lmt,Caceres:2018ehr,Leichenauer:2018tnq,Bilotta:2023hwq, Kawamoto:2025oko}. Apart from the holographic setup, an influential example within the four-dimensional standard model was presented by Maldacena, Milekhin, and Popov~\cite{Maldacena:2018gjk}. They demonstrated that traversable wormholes can be constructed without artificially introducing negative energy, but rather by exploiting vacuum energy contributions from fermions. The Casimir energy in their setup effectively takes the form of the energy–momentum tensor of bi-directional null dust, whose role in supporting wormhole geometries and dynamical formation have been investigated in Refs.~\cite{Hayward:2002pm,Koyama:2004uh,Koga:2025bqw}. Moreover, the formation of such wormholes through quantum tunneling is discussed in Ref.~\cite{Horowitz:2019hgb}. Finally, limitations arising from the quantum energy inequality have been analyzed in Ref.~\cite{Kontou:2024mtd}.

While the Maldacena--Milekhin--Popov construction is certainly striking in its own right, we would like to emphasize a different aspect of their work. Their construction is especially notable in that it treats the wormhole as a perturbation of the near-horizon geometry of a near-extremal black hole. This perturbative perspective highlights a new method of solution construction, one that can potentially be generalized to broader EFT settings. It is precisely this aspect that motivates our approach in the present work.

In this paper, we first reformulate the construction of wormhole solutions as perturbations around the near-horizon geometries of near-extremal Reissner--Nordstr\"{o}m and Myers--Perry black holes, abstracting the key features on the Maldacena--Milekhin--Popov construction~\cite{Maldacena:2018gjk}.
We then apply this construction to the EFT setup. However, we find that such perturbative wormhole solutions cannot be realized when the supporting energy is provided by higher derivative corrections in the effective field theory approach.
This conclusion holds independently of the details of the correction terms, since the enhanced symmetries of the near-horizon geometries impose strong constraints on the effective energy–momentum tensor near the throat. Our analysis therefore establishes no-go theorems: traversable wormholes cannot arise perturbatively from Reissner--Nordstr\"{o}m and Myers--Perry black holes in an effective field theory setup.

This paper is organized as follows.
In Sec.~\ref{sec: spherically symmetric wh}, we formulate the construction of perturbative wormhole solutions, focusing on the Reissner--Nordstr\"{o}m black hole background. Then, we apply this construction to EFT setup in Sec.~\ref{sec: no-go RN} and show a no-go theorem about charged wormholes.
In Sec.~\ref{sec:no go MP}, we generalize these consideration to the wormhole perturbed from an equal-angular-momenta Myers--Perry black hole, and we show a no-go theorem in a similar manner. The final section~\ref{sec:summary} is devoted to summary and discussion.

\section{Constructing traversable wormholes from near-horizon geometry}
\label{sec: spherically symmetric wh}
In this section, we review the strategy for constructing traversable wormhole solutions using the near-horizon geometry of black holes, as demonstrated in Ref.~\cite{Maldacena:2018gjk} for wormholes supported by the Casimir energy of a fermion localized on a magnetic flux. Here, we focus on the geometrical aspects of the construction and do not specify the physical origin of the negative energy sources that support the wormhole structure.
In this section, we restrict our attention to static, spherically symmetric black holes, while setups involving stationary black holes will be discussed in the later sections.

\subsection{Reissner--Nordstr\"{o}m black holes and near-horizon geometry}
In this section, we consider the Einstein--Maxwell system with some energy source in four dimensions, where the action is given by
\footnote{
We express the abstract tensors, including their basis, by boldface notation, basically following the convention in Ref.~\cite{Misner:1973prb}.
We also use the shorthand notion 
$\bm{d}x^{\mu} \bm{d}x^{\nu} \coloneqq \bm{d}x^{(\mu} \otimes \bm{d} x^{\nu)} = \frac{1}{2} (\bm{d}x^{\mu} \otimes \bm{d} x^{\nu} + \bm{d}x^{\nu} \otimes \bm{d} x^{\mu})$, while the wedge product is defined as $\bm{d} x^{\mu} \wedge \bm{d} x^{\nu} \coloneqq 2 \bm{d} x^{[\mu} \otimes \bm{d} x^{\nu]} = \bm{d} x^{\mu} \otimes \bm{d} x^{\nu} -  \bm{d} x^{\nu} \otimes \bm{d} x^{\mu} $.
}
\begin{align}
 S = \int d^4 x  \sqrt{-g} \left( \frac{1}{16 \pi G} R - \frac{1}{16 \pi k} F_{\mu\nu} F^{\mu\nu} \right) + \mathcal{S}, \label{Einstein-Maxwell action}
\end{align}
where $R$ is the Ricci scalar associated with the metric $\bm{g} = g_{\mu\nu} \bm{d}x^{\mu} \bm{d}x^{\nu}$ and $\bm{F} = \frac{1}{2} F_{\mu\nu} \bm{d}x^{\mu} \wedge \bm{d} x^{\nu}$ is the field strength of the electromagnetic field which is represented as the vector potential $\bm{A} = A_{\mu} \bm{d} x^{\mu}$ by
\begin{align}
 \bm{F} = \bm{d} \bm{A} = \partial_{[\mu} A_{\nu]} \bm{d} x^{\mu} \wedge \bm{d}x^{\nu}. 
\end{align}
The positive constants $G$ and $k$ are the Newton constant and the Coulomb constant, respectively. The term $\mathcal{S}$ in the action \eqref{Einstein-Maxwell action} represents contributions beyond the Einstein--Maxwell sector, which may consist of other dynamical degrees of freedom, vacuum energy, or higher derivative corrections, depending on the specific setup.

The equations of motion can be expressed  as
\begin{align}
E_{\mu\nu} \coloneqq \frac{16 \pi G}{\sqrt{-g}} \frac{\delta S}{\delta g^{\mu\nu}} &= R_{\mu\nu} - \frac{1}{2} g_{\mu\nu} R - 8 \pi G \left( T^{\text{EM}}_{\mu\nu} + \mathcal{T}_{\mu\nu} \right) = 0, \label{Einstein Eqs}\\
E^{\mu} \coloneqq \frac{4 \pi k}{\sqrt{-g}} \frac{\delta S}{\delta A_{\mu}} &= \nabla_{\nu} F^{\nu\mu} + 4 \pi k \mathcal{J}^{\mu} = 0, \label{Maxwell Eqs}
\end{align}
where the energy--momentum tensor of electromagnetic field can be expressed as 
\begin{align}
 T_{\mu\nu}^{\text{EM}} \coloneqq \frac{1}{4 \pi k} \left( F_{\mu\rho} F_{\nu}{}^{\rho} - \frac{1}{4} F_{\rho\sigma} F^{\rho\sigma} g_{\mu\nu} \right),
\end{align}
and  the contributions from $\mathcal{S}$ can be expressed as 
\begin{align}
 \mathcal{T}_{\mu\nu} &= - \frac{2}{\sqrt{-g}} \frac{\delta \mathcal{S}}{\delta g^{\mu\nu}},  \\
 \mathcal{J}^{\mu} &= \frac{1}{\sqrt{-g}} \frac{\delta \mathcal{S}}{\delta A^{\mu}}.
\end{align}
Similar to other geometric tensors, we use the expression
$\bm{\mathcal{T}} = \mathcal{T}_{\mu\nu} \bm{d}x^{\mu} \bm{d}x^{\nu}$ and $\bm{\mathcal{J}} = \mathcal{J}^{\mu} \bm{\partial}_{\mu}$.
Instead of using the vector potential, we regard the field strength $\bm{F}$ itself as an independent dynamical variable, with imposing the Bianchi identity
\begin{align}
 \bm{d} \bm{F} = \frac{1}{3!} ~ 3 \partial_{[\mu} F_{\nu\rho]} ~ \bm{d}x^{\mu} \wedge \bm{d}x^{\nu} \wedge \bm{d} x^{\nu} = 0.
\end{align}

Assuming $\mathcal{T}_{\mu\nu} = 0$ and $\mathcal{J}_{\mu} = 0$, for now, the static, spherically symmetric, asymptotically flat solution of the Einstein--Maxwell equations are expressed by the Reissner--Nordstr\"{o}m solutions,
\begin{align}
 \boldsymbol{g} &= - \mathrm{e}^{2 \psi(r)} f(r) \bm{d}t^2 + \frac{\bm{d}r^2}{f(r)} + r^2 \bm{d} \Omega^2, \notag\\
 \boldsymbol{F} &= - E_{r}(r) \mathrm{e}^{\psi(r)}\bm{d}t \wedge \bm{d}r + B_{r}(r) r^2 \sin\theta \bm{d}\theta \wedge \bm{d}\phi,
\end{align}
where functions $f(r), \psi(r), E_{r}(r)$ and $B_{r}(r)$ are given by 
\begin{align}
 f(r) &= 1 - \frac{2 G M}{r} + \frac{G (k Q_{e}^2 + k_{m} Q_{m}^2)}{r^2},\\
 \psi(r) &= 0, \\
 E_{r}(r) &=  k \frac{Q_{e}}{r^2}, \\
 B_{r}(r) &=  \frac{1}{4 \pi} \frac{Q_{m}}{r^2}.
\end{align}
Here we introduce magnetic Coulomb constant $k_{m}$ by
\begin{align}
 k_{m} \coloneqq \frac{1}{(4 \pi)^2 k}.
\end{align}

Note that the field strength corresponds to 
\begin{align}
 \boldsymbol{F} &= - k \frac{Q_{e}}{r^2} \bm{d}t \wedge \bm{d}r + \frac{1}{4 \pi} \frac{Q_{m}}{r^2} r^2 \sin\theta \bm{d}\theta \wedge \bm{d}\phi, \\
* \boldsymbol{F} &= \frac{1}{4 \pi} \frac{Q_{m}}{r^2} \bm{d}t \wedge \bm{d}r + k \frac{Q_{e}}{r^2} r^2 \sin\theta \bm{d}\theta \wedge \bm{d}\phi,
\end{align}
and hence the charges are defined so that
\begin{align}
 Q_{e} &= \frac{1}{4 \pi k} \int_{S^{2}} * \boldsymbol{F}, \\
 Q_{m} &= \int_{S^{2}} \boldsymbol{F}.
\end{align}

Let us consider the case where the Reissner--Nordstr\"{o}m solution possesses non-degenerate Killing horizons associated with the static Killing vector $ \boldsymbol{\xi}_{(t)} = \boldsymbol{\partial}_{t} $, that is, the case where $G M^2 >  k Q_{e}^2 + k_{m} Q_{m}^2 $ with $M > 0$. In this case, the outer Killing horizon is located at $r = r_{\text{H}}$ defined by
\begin{align}
r_{\text{H}} := G M + \sqrt{G} \sqrt{ G M^2 - \left( k Q_{e}^2 + k_{m} Q_{m}^2 \right)}
= G M \left( 1 + \epsilon \right),
\end{align}
with a parameter $0 < \epsilon \leq 1$ defined by
\begin{align}
 \epsilon := \sqrt{\frac{G M^2 - (k Q_{e}^2 + k_{m} Q_{m}^2)}{G M^2}}.
\end{align}
The limit $\epsilon \rightarrow 0$ corresponds to the extremal limit, and $\epsilon = 1$ corresponds to the Schwarzschild black hole. 
It is useful to use the parameters $r_{\text{H}}$ and $\epsilon$ instead of $M$ and $k Q_{e}^2 + k_{m} Q_{m}^2$, through the relation
\begin{align}
 G M = \frac{1}{1 + \epsilon} r_{\text{H}}, \\
 G (k Q_{e}^2 + k_{m} Q_{m}^2 ) =  \frac{1 - \epsilon}{1 + \epsilon} r_{\text{H}}^2. \label{Qe^2 + Qm^2 by epsilon rH}
\end{align}
In terms of $\epsilon$ and $r_{\text{H}}$, the function $f(r)$ can be expressed as 
\begin{align}
 f(r) = \frac{1}{r^2}(r - r_{\text{H}})\left(  r - \frac{1 - \epsilon}{1+\epsilon} r_{\text{H}}\right).
\end{align}
Then, the surface gravity $\kappa$ on the outer Killing horizon associated with $\boldsymbol{\xi}_{(t)}$, which is defined by $\nabla_{\rho}(g_{\mu\nu} \xi_{(t)}^{\mu} \xi_{(t)}^{\mu})|_{r = r_{\text{H}}} = - 2 \kappa g_{\rho\nu}\xi^{\nu}_{(t)}|_{r = r_{\text{H}}}$,  is given by
\begin{align}
 \kappa &= \frac{1}{2} f'(r_{\text{H}}) = \frac{1}{r_{\text{H}}} \frac{\epsilon}{1+\epsilon}.
\end{align}
The surface gravity $\kappa$ vanishes in the extremal limit $\epsilon \rightarrow 0$.

Let us introduce new coordinates $\tau_{\text{R}}, \rho_{\text{R}}$ by
\begin{align}
 \tau_{\text{R}} \coloneqq \kappa  t, \qquad \rho_{\text{R}} \coloneqq 1 + \frac{1}{\kappa} \frac{r - r_{\text{H}}}{r_{\text{H}}^2},
 \label{def R coordinates RN}
\end{align}
which can be inverted as,
\begin{align}
t = \frac{1}{\kappa} \tau_{\text{R}},\qquad r = r_{\text{H}} + \kappa~r_{\text{H}}^2 (\rho_{\text{R}} - 1).
\end{align}
The static spacetime region outside the outer horizon, given by $t \in (-\infty, \infty)$ and $r \in (r_{\text{H}}, \infty)$, corresponds to the coordinate region $\tau_{\text{R}} \in (- \infty, \infty)$ and $\rho_{\text{R}} \in (1, \infty)$.  Note that the subscript $\text{R}$ indicates that the coordinates $\{\tau_{\text{R}}, \rho_{\text{R}}\}$ can be interpreted as Rindler-like coordinates, as we will see below.
Using these coordinates, the Reissner--Nordstr\"{o}m spacetime can be described as
\begin{align}
\boldsymbol{g} = r_{\text{H}}^2   \left(
-   \frac{ \rho_{\text{R}}^2 - 1}{\left(1 + \delta(\rho_{\text{R}}) \right)^2} \bm{d}\tau_{\text{R}}^2
+ \frac{\left(1 + \delta(\rho_{\text{R}}) \right)^2}{\rho_{\text{R}}^2 - 1} \bm{d}\rho_{\text{R}}^2
+ \left(1 + \delta(\rho_{\text{R}}) \right)^2 \bm{d} \Omega^2
\right), \label{g Rindler}
\end{align}
where we introduced the function $\delta(\rho_{\text{R}})$ by
\begin{align}
 \delta(\rho_{\text{R}}) := \frac{r(\rho_{\text{R}}) - r_{\text{H}}}{r_{\text{H}}} = \kappa r_{\text{H}} (\rho_{\text{R}} - 1)= \frac{\epsilon}{1 + \epsilon}  (\rho_{\text{R}} - 1).
 \label{def delta RN}
\end{align}
The smallness of $\delta(\rho_{\text{R}})$ indicates how close the points at $\rho_{\text{R}}$ are to the outer Killing horizon.
The near-horizon region can be defined by
\begin{align}
 \delta(\rho_{\text{R}}) = \frac{r(\rho_{\text{R}}) - r_{\text{H}}}{r_{\text{H}}} \ll 1,
\end{align}
which can be translated into the condition in the $\rho_{\text{R}}$ coordinate as 
\begin{align}
  \rho_{\text{R}} - 1 \ll \frac{1 + \epsilon}{\epsilon}.
\end{align}
In the near-horizon region, $\delta$ terms in the metric components can be neglected, and the metric, therefore, can be expressed as 
\begin{align}
\boldsymbol{g} = r_{\text{H}}^2 \left(
-   (\rho_{\text{R}}^2 - 1) \bm{d}\tau_{\text{R}}^2
+ \frac{\bm{d}\rho_{\text{R}}^2}{ \rho_{\text{R}}^2 - 1 }
+ \bm{d} \Omega^2
\right) + \mathcal{O}(\delta), \label{gNH Rindler}
\end{align}
which is the metric of $\text{AdS}^{2} \times \text{S}^2$.

Then, the global AdS coordinates $\{\tau, \rho \}$ can be introduced by 
\begin{align}
 \tau &= \tan^{-1}\left( \sqrt{1 - \frac{1}{\rho_{\text{R}}^2}} \sinh \tau_{\text{R}} \right),
\\
 \rho &= \sqrt{\rho_{\text{R}}^2 - 1} \cosh \tau_{\text{R}},
\end{align}
which, inversely, can be expressed as  
\begin{align}
\tau_{\text{R}} &= \tanh^{-1} \left( \sqrt{1 + \frac{1}{\rho^2}} \sin \tau \right),\label{tau to Rindler}\\
 \rho_{\text{R}} &= \sqrt{1 + \rho^2} \cos \tau.\label{rho to Rindler}
\end{align}
We can express the near-horizon geometry $\text{AdS}^{2} \times \text{S}^2$ as
\begin{align}
 \boldsymbol{g} =
r_{\text{H}}^2
\left(
- (1 + \rho^2) \bm{d} \tau^2 + \frac{\bm{d}\rho^2}{1 + \rho^2} + \bm{d}\Omega^2\right) + \mathcal{O}(\delta)
.\label{gNH global}
\end{align}
The metric is well defined in the coordinate region given by $\tau \in (-\infty, \infty), \rho \in (- \infty, \infty)$, while the Rindler-like coordinates only cover the region $|\tau| < \rho$.
The horizon, $ \rho_{\text{R}} \to 1$ along $\tau_{\text{R}}$ constant line, corresponds to $\tau = \rho = 0$ in the global coordinates.
Note that the function $\delta$ is now expressed as
\begin{align}
 \delta(\rho, \tau) = \frac{\epsilon}{1 + \epsilon}\left( \sqrt{1 + \rho^2} \cos \tau - 1 \right).
\end{align}

In the near-horizon limit, the electromagnetic field strength can be evaluated as 
\begin{align}
  \boldsymbol{F} &= - k \frac{Q_{e}}{r^2} \bm{d}t \wedge \bm{d}r + \frac{1}{4 \pi} \frac{Q_{m}}{r^2} r^2 \sin\theta \bm{d}\theta \wedge \bm{d}\phi \notag \\
&= - k \frac{Q_{e}}{(1 + \delta(\rho_{\text{R}}))^2} \bm{d} \tau_{\text{R}} \wedge \bm{d} \rho_{\text{R}} + \frac{1}{4 \pi} Q_{m} \sin\theta \bm{d}\theta \wedge \bm{d}\phi \label{F Rindler} \\
&= \left( - k Q_{e} \bm{d} \tau_{\text{R}} \wedge \bm{d} \rho_{\text{R}} + \frac{1}{4 \pi} Q_{m} \sin\theta \bm{d}\theta \wedge \bm{d}\phi \right)  + \mathcal{O}(\delta) \notag \\
&= \left( - k Q_{e} \bm{d} \tau \wedge \bm{d} \rho + \frac{1}{4 \pi} Q_{m} \sin\theta \bm{d}\theta \wedge \bm{d}\phi \right) + \mathcal{O}(\delta). \label{FNH global} 
\end{align}
One can directly show that the metric \eqref{gNH global} of $\text{AdS}^2 \times \text{S}^{2}$ with the electromagnetic field strength \eqref{FNH global}, setting $\delta = 0$, is also an exact solution of Einstein--Maxwell equations.

\subsection{Wormhole as perturbations of near-horizon geometry}
\label{subsec:strategy RN}
Here we reformulate the strategy for constructing traversable wormhole by abstracting the study in Ref.~\cite{Maldacena:2018gjk} for the wormhole supported by the Casimir energy.
Let us consider static, spherically symmetric perturbation around near-horizon geometry of Reissner--Nordstr\"{o}m spacetime.
To treat the perturbations rigorously, we introduce dimensionless positive parameter $\lambda$ which counts the order of perturbations \cite{Wald:1984rg}.  
Thus, we consider one parameter family of static, spherically symmetric metric $\bm{g}(\lambda)$ as
\begin{align}
 \bm{g}(\lambda) = \ell(\lambda)^2 \left(
- \frac{ 1 + \rho^2}{\gamma(\lambda;\rho)} \bm{d}\tau^2 + \frac{ \gamma(\lambda; \rho)}{(1 + \rho^2)}\bm{d} \rho^2 
+ A(\lambda; \rho) \bm{d}\Omega^2
\right), \label{gWH full}
\end{align}
and the perturbations can be defined as a Taylor expansion with respect to $\lambda$,
\begin{align}
\bm{g}(\lambda) = \bar{\bm{g}} + \lambda \bm{g}^{(1)} + \frac{1}{2} \lambda^2 \bm{g}^{(2)} + \mathcal{O}(\lambda^3),
\end{align}
where
\begin{align}
 \bar{\bm{g}} \coloneqq \bm{g}(0),\qquad \bm{g}^{(1)} \coloneqq \left. \frac{d \bm{g}}{d \lambda} \right|_{\lambda = 0}, \qquad \bm{g}^{(2)}\coloneqq  \left.\frac{d^2 \bm{g}}{d \lambda^2}\right|_{\lambda=0}, 
\end{align}
and so on. We use similar notation for any quantities depending on $\lambda$.  Note that the function $A(\lambda;\rho)$ represents the area of $\tau, \rho$-constant two sphere normalized by $4 \pi \ell(\lambda)^2$. 
Using the gauge degrees of freedom of the coordinate choice, we include $\lambda$-dependence to the over all parameter $\ell$ as
\begin{align}
 \ell(\lambda) = \sqrt{G} \sqrt{k q_{e}(\lambda)^2 + k_{m} q_{m}(\lambda)^2},
\end{align}
where $q_{e}(\lambda)$ and $q_{m}(\lambda)$ are electric and magnetic charge of full spacetime appearing below.

Since we assume that the background is $\text{AdS}^2 \times S^2$, the functions $\gamma$ and $A$ can be expanded as 
\begin{align}
 \gamma(\lambda;\rho) &= 1 + \lambda \gamma^{(1)}(\rho) + \frac{1}{2} \lambda^2 \gamma^{(2)}(\rho) + \mathcal{O}(\lambda^2), \\
 A(\lambda;\rho) &= 1 + \lambda A^{(1)}(\rho) + \frac{1}{2} \lambda^2 A^{(2)}(\rho) + \mathcal{O}(\lambda^2),
\end{align}
and hence the background metric and metric perturbations can be expressed as 
\begin{align}
\bar{\boldsymbol{g}} & = \bar{\ell}^2
\left(
- (1 + \rho^2) \bm{d} \tau^2 + \frac{\bm{d}\rho^2}{1 + \rho^2} +\bm{d}\Omega^2  \right), \label{gWH0}\\
 \boldsymbol{g}^{(1)} & =  \bar{\ell}^2
\left(
\gamma^{(1)}(\rho) (1 + \rho^2) \bm{d} \tau^2 + \gamma^{(1)}(\rho) \frac{\bm{d}\rho^2}{1 + \rho^2}  + A^{(1)}(\rho) \bm{d}\Omega^2  \right) + 2 \bar{\ell} \ell^{(1)} \bar{\bm{g}}. \label{gWH1}
\end{align}

Similarly, the electromagnetic field strength is perturbed as 
\begin{align}
 \bm{F}(\lambda) = F_{\tau\rho}(\lambda;\rho) \bm{d}\tau \wedge \bm{d} \rho + F_{\theta\phi}(\lambda) \sin \theta \bm{d}\theta \wedge \bm{d} \phi, \label{FWH full}
\end{align}
where the electric part $F_{\tau \rho}$ can be expanded as 
\begin{align}
 F_{\tau \rho}(\lambda; \rho) = - k \bar{q}_{e} + \lambda F^{(1)}_{\tau\rho}(\rho) + \mathcal{O}(\lambda^2),
\end{align}
and the magnetic part $F_{\theta\phi}$, which must be constant in $\rho$ due to the Bianchi identity, can be expressed as 
\begin{align}
 F_{\theta \phi}(\lambda) = \frac{\bar{q}_{m}}{4 \pi} + \lambda \frac{q^{(1)}_{m}}{4 \pi} + \mathcal{O}(\lambda^2). \label{F theta phi solution}
\end{align}
Thus, the background electromagnetic field is given by
\begin{align}
\bar{\bm{F}} = - k \bar{q}_{e} \bm{d}\tau \wedge \bm{d} \rho + \frac{1}{4 \pi} \bar{q}_{m} \sin \theta \bm{d}\theta \wedge \bm{d} \phi. \label{FWH0}
\end{align}

We solve Einstein--Maxwell equations with assuming the source term $\bm{\mathcal{T}}$ and $\bm{\mathcal{J}}$ are at least first order in $\lambda$,
\begin{align}
 \bm{\mathcal{T}} &= \lambda \bm{\mathcal{T}}^{(1)} + \mathcal{O}(\lambda^2), \\
 \bm{\mathcal{J}} &= \lambda \bm{\mathcal{J}}^{(1)} + \mathcal{O}(\lambda^2).
\end{align}
Furthermore, we demand the following conditions for the perturbations, 
\begin{itemize}
 \item \textbf{Throat condition:}
There exists a throat $\rho = \rho_{\text{th}}$ which is defined by
\begin{align}
 \left. \partial_{\rho} A(\lambda; \rho) \right|_{\rho = \rho_{\text{th}}} = 0, \label{definition of throat}
\end{align}
and satisfies the flare-out condition 
\begin{align}
\left. \partial_{\rho}^2 A(\lambda; \rho) \right|_{\rho = \rho_{\text{th}}}  > 0. \label{throat condition}
\end{align}
 \item \textbf{Matching condition:} To obtain asymptotically flat solution, we join the wormhole geometry with the black hole geometry in the $\rho \gg 1$ region. 
For this matching, we require that, in the $\rho \to \infty$ limit keeping leading order correction terms, the metric reduces to the near-horizon ($\delta(\rho_{\text{R}}) \ll 1$) geometry of near-extremal ($\epsilon \ll 1$) black hole in the $\rho_{\text{R}} \gg 1$ region in the Rindler-like coordinates, up to linear order in $\delta$.
In the near-extremal limit, the function $\delta$ can be written as 
\begin{align}
 \delta(\rho_{\text{R}}) =  \epsilon \rho_{\text{R}} ( 1 + \mathcal{O}(\epsilon)).
\end{align}
Therefore, the approximation is consistent only when $1 \gg \rho_{\text{R}}^{-1} \gg \epsilon$. 
To perform the matching, we will omit terms of order $\mathcal{O}(\rho^{-1}_{\text{R}}, \epsilon)$, while keeping the terms linear in $\delta \approx \epsilon \rho_{\text{R}}$. Such a treatment is consistent when $\epsilon \rho_{\text{R}} \gg \mathcal{O}(\rho^{-1}_{\text{R}}, \epsilon)$.
Therefore, we eventually assume the following hierarchy in the matching region:
\begin{align}
 1 \gg \epsilon \rho_{\text{R}} \gg \frac{1}{\rho_{\text{R}}} \gg \epsilon. \label{hierarchy}
\end{align}
We require the same property for another side of wormhole, thus, for $ - \rho \gg 1$ region.
\end{itemize}

In the case where the outside black hole solution is given by the exact Reissner--Nordstr\"{o}m black hole, the metric and field strength in the Rindler-like coordinates, given by Eqs.~\eqref{g Rindler} and \eqref{F Rindler}, can be expressed as  
\begin{align}
 \bm{g}^{\text{RN}} &= r_{\text{H}}^2 \left( - \frac{1}{1 + 2 \epsilon \rho_{\text{R}}} \rho_{\text{R}}^2 \bm{d}\tau_{\text{R}}^2 + ( 1 + 2 \epsilon \rho_{\text{R}})  \frac{\bm{d} \rho_{\text{R}}^2}{\rho_{\text{R}}^2} + (1 + 2 \epsilon \rho_{\text{R}}) \bm{d}\Omega^2 \right) + \mathcal{O}( \epsilon, \rho_{\text{R}}^{-1}, (\epsilon \rho_{\text{R}})^2) , \label{gNH NE}\\
 \bm{F}^{\text{RN}} & =  - k Q_{e}(1 - 2 \epsilon \rho_{\text{R}}) \bm{d}\tau_{\text{R}} \wedge \bm{d} \rho_{\text{R}} + \frac{1}{4 \pi} Q_{m} \sin \theta \bm{d}\theta \wedge \bm{d}\phi + \mathcal{O}( \epsilon, \rho_{\text{R}}^{-1}, (\epsilon \rho_{\text{R}})^2). \label{FNH NE}
\end{align}
On the other hand, the wormhole solution, Eqs.~\eqref{gWH full} and \eqref{FWH full}, can be expressed in the $\rho \gg 1$ limit as 
\begin{align}
 \bm{g}(\lambda) &= \ell(\lambda)^2 \left(
- \frac{1}{\gamma(\lambda;\rho)}\rho^2 \bm{d}\tau^2 + \gamma(\lambda; \rho)\frac{\bm{d} \rho^2}{\rho^2} 
+ A(\lambda; \rho) \bm{d}\Omega^2
\right) + \mathcal{O}(\rho^{-2}), \label{gWH largerho}\\
 \bm{F}(\lambda) &= F_{\tau\rho}(\lambda;\rho) \bm{d}\tau \wedge \bm{d} \rho + F_{\theta\phi}(\lambda) \sin \theta \bm{d}\theta \wedge \bm{d} \phi. \label{FWH largerho}
\end{align}
Then, let us assume that the coordinates $(\tau, \rho)$ and $(\tau_{\text{R}}, \rho_{\text{R}})$ are related up to a constant scaling,
\begin{align}
 \tau = c_{1}(\lambda) \tau_{\text{R}}, \qquad
 \rho = c_{2}(\lambda) \rho_{\text{R}}.
\end{align}
Since we take the limit $1 \gg \rho_{\text{R}}^{-1}$ and $1 \gg \rho^{-1}$ simultaneously, parameter $c_{2}$ must satisfy,
\begin{align}
 c_{2} \gg \frac{1}{\rho_{\text{R}}} \gg \epsilon, \label{hierarchy c2}
\end{align}
where the last relation comes from the relation \eqref{hierarchy}. We regard $c_{1}, c_{2}$ as $\mathcal{O}(1)$  quantities in the approximation.
Then, comparing Eqs.~\eqref{gNH NE} and \eqref{FNH NE} with Eqs.~\eqref{gWH largerho} and \eqref{FWH largerho}, respectively, we find that $\gamma, A$ and $F_{\tau\rho}$ must be linear in $\rho$ in the limit $\rho \to \infty$,
\begin{align}
 \gamma(\lambda; \rho) &= c_{3}(\lambda)^2 \left(  1 + c_{4}(\lambda) \rho \right) + \mathcal{O}(c_{4}, \rho^{-1}, (c_{4} \rho)^2), \label{matching gamma} \\ 
 A(\lambda; \rho) & =  c_{3}(\lambda)^2 \left(  1 + c_{4}(\lambda) \rho \right) + \mathcal{O}(c_{4}, \rho^{-1}, (c_{4} \rho)^2), \label{matching A} \\
 F_{\tau\rho}(\lambda;\rho) &= c_{5}(\lambda)(1 - c_{4}(\lambda) \rho) + \mathcal{O}(c_{4}, \rho^{-1}, (c_{4} \rho)^2) , \label{matching F}
\end{align}
with identifying the parameters $c_{i}$ with those of a black hole through
\begin{align}
 r_{\text{H}} &= \ell c_{3} = \ell \frac{c_{1} c_{2}}{c_{3}}, \label{matching rH}\\
 \epsilon &= \frac{1}{2} c_{4} c_{2}, \label{matching epsilon}\\
 Q_{e} &=  - \frac{c_{5} c_{1} c_{2}}{k} = - \frac{c_{5} c_{3}^2}{k}, \label{matching Qe}  \\
Q_{m} &= 4 \pi F_{\theta \phi}(\lambda) . \label{matching Qm}
\end{align}
Note that the parameter $c_{1}$ must be introduced so that 
\begin{align}
 c_{1} = c_{3}^2/c_{2},
\end{align}
while the constant $c_{2}$ leaves unfixed.

Note that the remaining free parameter $c_{2}$ represents a possible scale difference between the natural coordinates used for the near throat wormhole geometry and the asymptotic black hole geometry. However, this does not affect the relation between the wormhole coordinates $(\tau, \rho)$ and the asymptotic coordinates $(t,r)$ in the leading order approximation, which can be  derived from Eqs.~\eqref{def R coordinates RN} and \eqref{def delta RN}, 
\begin{align}
 t = \frac{1}{\kappa} \tau_{\text{R}} = \frac{1}{\kappa c_{1}} \tau = \frac{c_{2}}{\kappa c_{3}^2} \tau = \frac{2 r_{\text{H}}}{c_{3}^2 c_{4}} \tau + \mathcal{O}(\epsilon) \label{t by tau},
\end{align}
and 
\begin{align}
 \frac{r - r_{\text{H}}}{r_{\text{H}}} = \kappa r_{\text{H}}\rho_{\text{R}} \left( 1 + \mathcal{O}(\rho_{\text{R}}^{-1}) \right) = \frac{c_{4}}{2} \rho \left( 1 + \mathcal{O}(\epsilon, c_{2} \rho_{\text{R}}^{-1}) \right).
\end{align}
Thus, it should be determined by the matching condition in more higher-order perturbations. Note also that the coefficient in Eq.~\eqref{t by tau} is nothing but the throat length parameter $\ell$ in Ref.~\cite{Maldacena:2018gjk}, which hereafter we express $\ell^{\text{MMP}}$ for avoiding notational conflict. Thus,
\begin{align}
 \ell^{\text{MMP}} = \frac{2 r_{\text{H}}}{c_{3}^2 c_{4}}.
\end{align}

We would like to emphasize that the conditions \eqref{matching gamma} - \eqref{matching F} are for matching to the exact Reissner--Nordstr\"{o}m black hole. 
More generally, it is possible that the contribution from the energy--momentum tensor is not negligible in the outside black hole region. In such a case, outside black hole geometry is corrected by $\mathcal{O}(\lambda)$ term.  
In the dominant contributions of the near-horizon expansion, such corrections could affect the constant terms. Because of this, the constants in the matching conditions are possibly modified and hence, they are, in general, given by 
\begin{align}
 \gamma(\lambda; \rho) & \approx c_{3,\gamma}(\lambda)^2 \left(  1 + c_{4,\gamma}(\lambda) \rho \right) 
 \\ 
 A(\lambda; \rho) & \approx  c_{3,A}(\lambda)^2 \left(  1 + c_{4,A}(\lambda) \rho \right)
 \\
 F_{\tau\rho}(\lambda;\rho) &\approx c_{5}(\lambda)(1 - c_{4,F}(\lambda) \rho),
\end{align}
in the $\rho \to \infty$ limit.
The explicit relation among the parameters relies on the explicit black hole solution with source term and hence depends on the specific setup.  
Although we focused only on the $\rho \to \infty$ limit here, we require similar conditions in the $\rho \to - \infty$ limit.

The purpose of this paper is to clarify under what conditions the above wormhole construction is possible.  In particular, we establish a no-go theorem showing that such a construction fails if only self-interactions of $\bm{g}$ and $\bm{F}$ are included as corrections to the Einstein–Maxwell action. Before discussing the no-go theorem, let us briefly consider an explicit example where the construction works below.

\subsection{Example: traversable wormholes with Casimir energy}
\label{sec:Casimir wormhole}
A setup which realizes the above construction of wormhole solution is studied in Ref.~\cite{Maldacena:2018gjk}, where the following energy--momentum tensor is considered as a trace free contributions of the Casimir energy of a fermion,
\begin{align}
\bm{\mathcal{T}} &= \mathcal{T}_{\mu\nu} \bm{d} x^{\mu} \bm{d}x^{\nu} = \lambda \mathcal{T}^{(1)} (\rho) \left( (1 +\rho^2) \bm{d} \tau^2 + \frac{\bm{d} \rho^2}{1 + \rho^2} \right) + \mathcal{O}(\lambda^2),\label{T Casimir} \\
\bm{\mathcal{J}} &= 0.
\end{align}
The energy conservation requires
\begin{align}
 (\nabla_{\mu} T^{\mu\nu}) \bm{\partial}_{\nu} = \frac{1+ \rho^2}{\bar{\ell}^4} \left( \partial_{\rho} \mathcal{T}^{(1)} + \frac{2 \rho}{1 + \rho^2} \mathcal{T}^{(1)}(\rho) \right)\bm{\partial}_{\rho} = 0,
\end{align}
which can be solved as 
\begin{align}
\mathcal{T}^{(1)}(\rho) =  \frac{\mathcal{E}}{1 + \rho^2},
 \end{align}
with an integration constant $\mathcal{E}$ which characterizes the energy density.

Then, the Maxwell equations~\eqref{Maxwell Eqs} at the first order in perturbations are given by 
\begin{align}
 E^{(1)}{}^{\mu} \bm{\partial}_{\mu} = \frac{1}{\bar{\ell}^4} \left( \partial_{\rho}F^{(1)}_{\tau\rho}(\rho) - k \bar{q}_{e} \partial_{\rho} A^{(1)}(\rho)\right) \bm{\partial}_{\tau} = 0,
\end{align}
which can be solved as 
\begin{align}
 F^{(1)}_{\tau \rho} = - k q^{(1)}_{e} + k \bar{q}_{e} A^{(1)}(\rho), \label{delta F tau rho solution}
\end{align}
with integration constant $q^{(1)}_{e}$.

One combination of the Einstein equations can be expressed as 
\begin{align}
 E^{(1)}{}^{\tau}{}_{\tau} - E^{(1)}{}^{\rho}{}_{\rho} &= \frac{1 + \rho^2}{\bar{\ell}^2} \partial_{\rho}^2 A^{(1)} - 8 \pi G \left( \mathcal{T}^{(1)}{}^{\tau}{}_{\tau} - \mathcal{T}^{(1)}{}^{\rho}{}_{\rho} \right) \label{Ein throat general}\\
&= \frac{1+ \rho^2}{\bar{\ell}^2} \left( \partial_{\rho}^2 A^{(1)} + \frac{16 \pi G \mathcal{E}}{(1 + \rho^2)^2} \right) = 0. \label{Ein throat Casimir}
\end{align}
An immediate consequence of this equation is that
\begin{align}
 \partial_{\rho}^2 A( \lambda; \rho)  &= \lambda \partial_{\rho}^2 A^{(1)} + \mathcal{O}(\lambda^2)\\
&= - \frac{16 \pi G \lambda \mathcal{E}}{(1 + \rho^2)^2}  + \mathcal{O}(\lambda^2),
\end{align}
and hence the throat condition can be satisfied for the negative energy, $\mathcal{E} < 0$.

By integrating the equation \eqref{Ein throat Casimir}, we obtain
\begin{align}
 A^{(1)} = a_{1} + a_{2} \rho - 8 \pi G \mathcal{E} \rho \arctan \rho, \label{delta A Cas by a1 a2}
\end{align}
where $a_{1}$ and $a_{2}$ are integration constants.

Let us focus on a simple class of solution which satisfies symmetric condition
\begin{align}
 \gamma(\lambda; - \rho) = \gamma(\lambda; \rho),\qquad A(\lambda; - \rho) = A(\lambda; \rho). \label{symmetric condition}
\end{align}
Thus, we focus on the case $a_{2} = 0$, and the throat is located at $\rho = 0$.
The other constant $a_{1}$ can be determined by plugging the expression \eqref{delta A Cas by a1 a2} into $E^{(1)}{}^{\rho}{}_{\rho} = 0$, which reduces to 
\begin{align}
 a_{1} = - 8 \pi G \mathcal{E}.
\end{align}
By plugging above expressions into the $E^{(1)}{}^{\theta}{}_{\theta} = 0$, we obtain
\begin{align}
 E^{(1)}{}^{\theta}{}_{\theta} = - \frac{1 + \rho^2}{2 \bar{\ell}^{2}} \left( \partial_{\rho}^2 \gamma^{(1)} + \frac{4 \rho}{1 + \rho^2} \partial_{\rho} \gamma^{(1)} + \frac{2}{1 + \rho^2} \gamma^{(1)} + 4 8 \pi G \mathcal{E} \frac{1 + \rho \arctan \rho}{1 + \rho^2}\right) = 0,
\end{align}
and this equation can be solved as
\begin{align}
 \gamma^{(1)}(\rho) = \frac{b_{1}}{1 + \rho^2} + \frac{b_2 \rho}{1 + \rho^2} - 8 \pi G \mathcal{E} \frac{\rho^2 + (3 + \rho^2) \rho \arctan \rho - \log(1 + \rho^2)}{1 + \rho^2}.
\end{align}
From the symmetric condition, we obtain $b_{2} = 0$.

Furthermore, in the $\rho \gg 1$ limit, we obtain
\begin{align}
 A(\lambda; \rho) & \approx  (1 - 8 \pi G \lambda \mathcal{E} )  - 4 \pi^2 G \lambda \mathcal{E} \rho  + \mathcal{O}(\lambda^2), \\
 \gamma(\lambda; \rho) & \approx (1 - 8\pi G \lambda \mathcal{E}) - 4 \pi^2 G \lambda \mathcal{E} \rho + \mathcal{O}(\lambda^2), \\
 F_{\tau\rho} (\lambda; \rho) & \approx - k (\bar{q}_{e} + \lambda q^{(1)}_{e} + 8 \pi G \lambda \mathcal{E} \bar{q}_{e} )  - 4 \pi^2 G \lambda \mathcal{E} k \bar{q}_{e} \rho   + \mathcal{O}(\lambda^2).
\end{align}
Hence the matching conditions with the exact Reissner--Nordstr\"{o}m black hole, Eqs.~\eqref{matching gamma}~-~\eqref{matching F} are satisfied with identifying the coefficients $c_{i}$ as  
\begin{align}
 c_{3}^2 &= 1 - 8 \pi G \lambda \mathcal{E} + \mathcal{O}(\lambda^2), \\
 c_{4} &= - 4 \pi^2 G \lambda \mathcal{E} + \mathcal{O}(\lambda^2), \\
 c_{5} &= - k (\bar{q}_{e} + \lambda q^{(1)}_{e} + 8 \pi G \lambda \mathcal{E} \bar{q}_{e}) + \mathcal{O}(\lambda^2).
\end{align}
Since $c_{4}$ is assumed to be $c_{4} \ll 1$ in order to match the near-horizon geometry of Reissner--Nordstr\"{o}m black hole, we obtain
\begin{align}
 c_{3}^2 &= 1 + \mathcal{O}(c_{4}, \lambda^2), \\
 c_{5} &= - k q_{e}(\lambda) + \mathcal{O}(c_{4}, \lambda^2).
\end{align} 

To summarize, in the system where the energy--momentum tensor can be given in the form of Eq.~\eqref{T Casimir}, one can construct wormhole solution when $\mathcal{E} < 0$ as perturbations from the near-horizon geometry of near-extremal black hole.
From the matching condition, the parameters for the matched black hole can be obtained as 
\begin{align}
 r_{\text{H}} &= \sqrt{G (k q_{e}(\lambda)^2 + k_{m} q_{m}(\lambda)^2) } + \mathcal{O}(\epsilon, \lambda^2), \\ 
Q_{e} &= q_{e}(\lambda) + \mathcal{O}(\epsilon, \lambda^2), \\
Q_{m} &= q_{m}(\lambda),
\end{align}
and 
\begin{align}
 \epsilon &= - 2 \pi^2 G \lambda \mathcal{E} c_{2} + \mathcal{O}(\lambda^2).
\end{align}

Note that the throat length parameter $\ell^{\text{MMP}}$ is given by
\begin{align}
 \ell^{\text{MMP}} = \frac{4 r_{\text{H}}}{\pi (- 8\pi G \lambda \mathcal{E})}, 
\end{align}
which reproduce the result in Ref.~\cite{Maldacena:2018gjk}.

\section{No-go theorems for wormholes perturbed from Reissner--Nordstr\"{o}m black holes}
\label{sec: no-go RN}
In this section, we show that the construction of wormhole solutions explained in the previous section fails in the general setup where the corrections to the Einstein--Maxwell action are limited to the self-interaction and treat such corrections perturbatively. 

\subsection{Higher derivative corrections to the Einstein--Maxwell action}
In this section, we focus on the system where the action is given by Eq.~\eqref{Einstein-Maxwell action}, with the correction term $\mathcal{S}$ is covariantly constructed from the metric $\bm{g}$ and the electromagnetic field $\bm{F}$, and their derivatives.  
Schematically, we express it as $\mathcal{S}[\bm{g}, \bm{F}]$.
We solve the Einstein--Maxwell equations perturbatively following the strategy in the previous section.
Since we treat the correction term as a source for the first-order perturbations, we assign the expansion parameter $\lambda$ to the action functional as 
 \begin{align}
  \mathcal{S}[\lambda; \bm{g}, \bm{F}] = \lambda \mathcal{S}^{(1)}[\bm{g}, \bm{F}] + \mathcal{O}(\lambda^2).
 \end{align}
Hence, the Einstein--Maxwell Equations for full variables can be expressed as
\begin{align}
 E^{\mu}{}_{\nu}(\lambda) &= R^{\mu}{}_{\nu}(\lambda) - \frac{1}{2} \delta^{\mu}{}_{\nu} R(\lambda) - 8 \pi G \left( T^{\text{EM}}{}^{\mu}{}_{\nu}(\lambda) + \mathcal{T}^{\mu}{}_{\nu}(\lambda; \bm{g}(\lambda), \bm{F}(\lambda) )\right) = 0, \\
E^{\mu}(\lambda) &= \nabla_{\nu} F^{\nu\mu} (\lambda) - \mathcal{J}^{\mu}(\lambda; \bm{g}(\lambda), \bm{F}(\lambda)) = 0.
\end{align}

By extracting the coefficients of the Taylor series of $\lambda$, we obtain the equations of motion for each order of perturbations.
For the background, we obtain sourceless Einstein--Maxwell equations and consider the near-horizon geometry $\text{AdS}^{2} \times \text{S}^{2}$ of Reissner--Nordstr\"{o}m black hole, Eqs.~\eqref{gWH0} and \eqref{FWH0}, as the background solutions. 
Then, the equations of motion for the first-order perturbations can be obtained as 
\begin{align}
  &R^{(1)}{}^{\mu}{}_{\nu} - \frac{1}{2} \delta^{\mu}{}_{\nu} R^{(1)} - 8 \pi G \left( T^{(1)}{}^{\text{EM}}{}^{\mu}{}_{\nu} + \mathcal{T}^{(1)}{}^{\mu}{}_{\nu}(\bar{\bm{g}}, \bar{\bm{F}} )\right) = 0, \\
&(\nabla_{\nu} F^{\nu\mu})^{(1)} - \mathcal{J}{}^{(1)}{}^{\mu}(\bar{\bm{g}}, \bar{\bm{F}}) = 0.
\end{align}

An example of the correction term is the effective action after integrating out the charged massive fermion in the quantum electrodynamics setup, thus, Euler--Heisenberg~\cite{Heisenberg:1936nmg} or Drummond--Hathrel action~\cite{Drummond:1979pp}, assigning expansion parameter $\lambda$ for each coupling constants,
\begin{align}
 \mathcal{S}[\lambda;\bm{g}, \bm{F}] =  \int d^4 x \sqrt{-g} \Biggl[&
\lambda \alpha_{1} (F_{\mu\nu} F^{\mu\nu})^2 + \lambda \alpha_{2} F_{\mu}{}^{\nu}F_{\nu}{}^{\rho}F_{\rho}{}^{\sigma}F_{\sigma}{}^{\mu} \notag\\
& + \lambda \alpha_{3} R F_{\mu\nu} F^{\mu\nu} + \lambda \alpha_{4} R_{\mu\nu} F^{\mu\rho}F^{\nu}{}_{\rho} + \lambda \alpha_{5} R_{\mu\nu\rho\sigma} F^{\mu\nu} F^{\rho\sigma} \notag\\
& + \lambda \alpha_{6} R^2 + \lambda \alpha_{7} R_{\mu\nu} R^{\mu\nu} + \lambda \alpha_{8} R_{\mu\nu\rho\sigma} R^{\mu\nu\rho\sigma}
\Biggr] + \mathcal{O}(\lambda^2),
\end{align}
although, our discussion below does not depend on this concrete setup.

\subsection{No-go theorem for static charged wormholes at first order in perturbations}
The key insight to show the no-go theorem for perturbative wormhole construction is that the possible form of the energy--momentum tensor $\bm{\mathcal{T}}^{(1)}(\bar{\bm{g}}, \bar{\bm{F}})$ and the electric flux $\bm{\mathcal{J}}^{(1)}(\bar{\bm{g}}, \bar{\bm{F}})$ are strongly restricted from the enhanced symmetry of near-horizon geometry;
The background metric $\bar{\bm{g}}$, given by Eq.~\eqref{gWH0}, possesses the Killing vectors about the symmetry of $\text{AdS}^{2}$, that is, $\text{SO}(2,1)$,  
\begin{align}
 \boldsymbol{\xi}_{1} &= \frac{\rho}{\sqrt{1 + \rho^2}} \cos \tau  \boldsymbol{\partial}_{\tau} + \sqrt{1 + \rho^2} \sin \tau \boldsymbol{\partial}_{\rho} ,\label{xi1} \\
 \boldsymbol{\xi}_{2} &= \frac{\rho}{\sqrt{1 + \rho^2}} \sin \tau  \boldsymbol{\partial}_{\tau} - \sqrt{1 + \rho^2} \cos \tau \boldsymbol{\partial}_{\rho}, \label{xi2} \\
 \boldsymbol{\xi}_{3} &= \boldsymbol{\partial}_{\tau}, \label{xi3}
\end{align}
in addition to those for the spherical symmetry 
\begin{align}
 \bm{J}_{1} &= \sin \phi \bm{\partial}_{\theta} + \frac{\cos \phi}{\tan \theta} \bm{\partial}_{\phi}, \label{J1}\\
 \bm{J}_{2} &=  - \cos \phi \bm{\partial}_{\theta} + \frac{\sin \phi}{\tan \theta} \bm{\partial}_{\phi}, \label{J2}\\ 
 \bm{J}_{3} &=  \bm{\partial}_{\phi}. \label{J3}
\end{align}
In other words, the background metric $\bar{\bm{g}}$ satisfies
\begin{align}
 \mathsterling_{\bm{\xi}_{i}} \bar{\bm{g}} = 0, \qquad  \mathsterling_{\bm{J}_{i}} \bar{\bm{g}} = 0,
\end{align}
for $i = 1, 2, 3$. Also, the electromagnetic field possesses the same symmetries,
\begin{align}
 \mathsterling_{\bm{\xi}_{i}} \bar{\bm{F}} = 0, \qquad  \mathsterling_{\bm{J}_{i}} \bar{\bm{F}} = 0.
\end{align}
Since the source terms in the first order equations of motion, $\bm{\mathcal{T}}^{(1)}(\bar{\bm{g}}, \bar{\bm{F}})$ and $\bm{\mathcal{J}}^{(1)}(\bar{\bm{g}}, \bar{\bm{F}})$, are covariantly constructed from $\bar{\bm{g}}$, $\bar{\bm{F}}$, and their derivatives, they respect the same symmetries as the metric and electromagnetic field strength, thus, 
\begin{align}
  \mathsterling_{\bm{\xi}_{i}} \bm{\mathcal{T}}^{(1)}(\bar{\bm{g}}, \bar{\bm{F}}) = 0, \qquad  \mathsterling_{\bm{J}_{i}} \bm{\mathcal{T}}^{(1)}(\bar{\bm{g}}, \bar{\bm{F}}) = 0, \label{LieD of T}
\end{align}
and
\begin{align}
  \mathsterling_{\bm{\xi}_{i}} \bm{\mathcal{J}}^{(1)}(\bar{\bm{g}}, \bar{\bm{F}}) = 0, \qquad  \mathsterling_{\bm{J}_{i}} \bm{\mathcal{J}}^{(1)}(\bar{\bm{g}}, \bar{\bm{F}}) = 0. \label{LieD of J}
\end{align}

Let us derive the general form of $\bm{\mathcal{T}}$ that satisfies Eqs.~\eqref{LieD of T}.
First, the symmetry about $\boldsymbol{\xi}_{3}$ indicates that the components of $\bm{\mathcal{T}}^{(1)}(\bar{\bm{g}}, \bar{\bm{F}})$ in the coordinates $\{\tau, \rho, \theta, \phi\}$ do not depend on $\tau$. Combining the spherical symmetries $\boldsymbol{J}_{i}$, the energy--momentum tensor can be generally expressed as 
\begin{align}
 \boldsymbol{\mathcal{T}}^{(1)}(\bar{\bm{g}}, \bar{\bm{F}})  = \tau_{\tau\tau}(\rho) (1 + \rho^2) \bm{d}\tau^2 + \tau_{\tau\rho}(\rho) (\bm{d}\tau \bm{d}\rho + \bm{d} \rho \bm{d} \tau) + \frac{\tau_{\rho\rho}(\rho)}{1 + \rho^2} \bm{d} \rho^2 + \tau_{\theta\theta}(\rho) \bm{d}\Omega^2.
\end{align}
Then, the Lie derivative along $\boldsymbol{\xi}_{1}$ can be evaluated as 
\begin{align}
  \mathsterling_{\boldsymbol{\xi}_{1}} \boldsymbol{\mathcal{T}}^{(1)}(\bar{\bm{g}}, \bar{\bm{F}})
&= \left(  \frac{2 \cos \tau}{\sqrt{1 + \rho^2}} \tau_{\tau \rho}(\rho)  + \sqrt{1 + \rho^2}  \sin \tau  ~\partial_{\rho} \tau_{\tau\tau}(\rho) \right) (1 + \rho^2)  \bm{d}\tau^2 \notag\\
&\qquad  + \left(  \frac{\cos \tau}{\sqrt{1 + \rho^2}} ( \tau_{\tau\tau}(\rho) + \tau_{\rho\rho}(\rho)) + \sqrt{1 + \rho^2} \sin \tau~\partial_{\rho}\tau_{\tau\rho}(\rho) \right) \left( \bm{d}\tau \bm{d}\rho + \bm{d}\rho \bm{d} \tau \right) \notag\\
&\qquad  + \left( \frac{2 \cos \tau}{\sqrt{1 + \rho^2}} \tau_{\tau\rho}(\rho) + \sqrt{1 + \rho^2} \sin \tau~ \partial_{\rho}\tau_{\rho\rho}(\rho) \right) \frac{\bm{d}\rho^2}{1 + \rho^2}\notag\\
&\qquad  + \sqrt{1 + \rho^2} \sin \tau ~\partial_{\rho} \tau_{\theta\theta}(\rho) \bm{d} \Omega^2.
\end{align}
The condition $\mathsterling_{\bm{\xi}_{1}} \bm{\mathcal{T}}^{(1)} = 0$ can be satisfied only when
\begin{align}
\tau_{\tau\tau} = \text{const.},  ~ \tau_{\rho\rho} = \text{const.}, ~\tau_{\theta\theta} = \text{const.}, ~ \tau_{\tau\rho} = 0,
\end{align}
with
\begin{align}
 \tau_{\tau\tau} + \tau_{\rho\rho} = 0. \label{sumtau=0}
\end{align}
Under these conditions, the remaining condition $ \mathsterling_{\boldsymbol{\xi}_{2}} \boldsymbol{\mathcal{T}}^{(1)}(\bar{\bm{g}}, \bar{\bm{F}}) = 0$ is automatically satisfied.  Thus, generally, the energy--momentum tensor can be expressed by two parameters $\mathcal{E}$ and $\mathcal{P}$ as 
\begin{align}
 \boldsymbol{\mathcal{T}}^{(1)}(\bar{\bm{g}}, \bar{\bm{F}}) =  \mathcal{E} \left( (1 + \rho^2) \bm{d}\tau^2 - \frac{\bm{d}\rho^2}{1 + \rho^2} \right) + \mathcal{P} \bm{d}\Omega^2.\label{delta T general}
\end{align} 
Note that the relative sign of $\tau$-$\tau$ and $\rho$-$\rho$ components are opposite to the case \eqref{T Casimir}.

Similarly, from the time translation symmetry $\bm{\xi}_{3}$ and the spherical symmetries $\bm{J}_{i}$, the electric current $\bm{\mathcal{J}}^{(1)}(\bar{\bm{g}}, \bar{\bm{F}})$ can be expressed as 
\begin{align}
 \bm{\mathcal{J}}^{(1)}(\bar{\bm{g}}, \bar{\bm{F}}) = \mathcal{J}^{(1)}{}^{\tau}(\rho) \bm{\partial}_{\tau} + \mathcal{J}^{(1)}{}^{\rho}(\rho) \bm{\partial}_{\rho}.
\end{align}
Then, the condition $\mathsterling_{\bm{\xi}_{1}} \bm{\mathcal{J}}^{(1)} = 0$ can be expressed as 
\begin{align}
 \mathsterling_{\bm{\xi}_{1}} \bm{\mathcal{J}}^{(1)}(\bar{\bm{g}}, \bar{\bm{F}})
&=
\left( - \mathcal{J}^{(1)}{}^{\rho}(\rho) \frac{\cos \tau}{(1 + \rho^2)^{3/2}} + \left( \frac{\rho}{\sqrt{1 + \rho^2}} \mathcal{J}^{(1)}{}^{\tau}(\rho)+ \sqrt{1 + \rho^2} \partial_{\rho} \mathcal{J}^{(1)}{}^{\rho}(\rho)  \right)\sin \tau    \right) \bm{\partial}_{\tau} \notag \\
&\qquad + \left( - \sqrt{1 + \rho^2} \mathcal{J}^{(1)}{}^{\tau}(\rho) \cos \tau + \left( - \frac{\rho}{\sqrt{1 + \rho^2}} \mathcal{J}^{(1)}{}^{\rho}(\rho) + \sqrt{1 + \rho^2} \partial_{\rho} \mathcal{J}^{(1)}{}^{\rho}(\rho) \right) \sin \tau \right) \bm{\partial}_{\rho} \notag\\
& = 0,
\end{align}
which can be satisfied only when $\mathcal{J}^{(1)}{}^{\tau} = \mathcal{J}^{(1)}{}^{\rho} = 0$. Thus, from the symmetry \eqref{LieD of J}, we obtain 
\begin{align}
 \bm{\mathcal{J}}^{(1)}(\bar{\bm{g}}, \bar{\bm{F}}) = 0.
\end{align}

Now, we are ready to show the no-go theorem for constructing wormhole solutions as perturbations from the near-horizon geometry of Reissner--Nordstr\"{o}m black hole.
The key property follows from the expression \eqref{delta T general} is that
\begin{align}
 \mathcal{T}^{(1)}{}^{\tau}{}_{\tau} = - \frac{\mathcal{E}}{\bar{\ell}^2},  \qquad \mathcal{T}^{(1)}{}^{\rho}{}_{\rho} = - \frac{\mathcal{E}}{\bar{\ell^2}},
\end{align}
where indices are lifted by the background metric $\bar{g}^{\mu\nu}$.
Hence, the contributions from the energy--momentum tensor in the Einstein equations for ${}^\tau{}_{\tau} - {}^{\rho}{}_{\rho}$ component,  Eq.~\eqref{Ein throat general}, are canceled, and the equation now can be evaluated as 
\begin{align}
0 &=  E^{(1)}{}^{\tau}{}_{\tau} -  E^{(1)}{}^{\rho}{}_{\rho} \notag\\
&= \frac{1 + \rho^2}{\bar{\ell}^2} \partial_{\rho}^2 A^{(1)}(\rho) - 8 \pi G \left( \mathcal{T}^{(1)}{}^{\tau}{}_{\tau} - \mathcal{T}^{(1)}{}^{\rho}{}_{\rho} \right) \notag\\
 &= \frac{1 + \rho^2}{\bar{\ell}^2} \partial_{\rho}^2 A^{(1)}(\rho).
\end{align}
Hence we obtain
\begin{align}
 \partial_{\rho}^2 A^{(1)}(\rho) = 0.\label{no throat condition}
\end{align}
Thus, including the background, we obtain
\begin{align}
 \partial_{\rho}^2 A(\lambda;\rho) = \mathcal{O}(\lambda^2),
\label{no throat condition up to first order}
\end{align}
and the flare-out condition \eqref{throat condition} cannot be satisfied for any value of $\rho_{\text{th}}$ at the first-order perturbations. 
This proves a no-go theorem for construction of wormhole solutions as the first-order perturbation from the near-horizon geometry of Reissner--Nordstr\"{o}m solution.

\subsection{No-go theorem for static charged wormholes at higher order in perturbations}
\label{subsec:no go RN higher}
The result in the previous subsection shows that the energy--momentum tensor respecting the symmetry of $\text{AdS}^{2} \times \text{S}^{2}$ cannot be a source for supporting the wormhole throat structure at the first order in perturbations.
However, the flare-out condition is unsatisfied marginally \eqref{no throat condition up to first order} and it is possibly satisfied by the second-order perturbations $\mathcal{O}(\lambda^2)$. Let us proceed our consideration to second-order perturbations, where the first-order perturbations possibly become a energy--momentum source without respecting the symmetry. 

To investigate it, let us consider general solutions for the first-order perturbations which satisfy one of the throat condition \eqref{definition of throat} and the matching conditions \eqref{matching A} at the first-order perturbations. 
Since the Maxwell equation is same as the Casimir energy case studied in Sec.~\ref{sec:Casimir wormhole}, the electromagnetic field can be given by Eqs.~\eqref{delta F tau rho solution} and \eqref{F theta phi solution}, as 
\begin{align}
 F^{(1)}_{\tau\rho} &= - k q^{(1)}_{e} + k \bar{q}_{e} A^{(1)}(\rho), \\ 
 F^{(1)}_{\theta\phi} &= \frac{1}{4 \pi} q^{(1)}_{m}.
\end{align}

As demonstrated above, difference of ${}^\tau{}_{\tau}$ and ${}^{\rho}{}_{\rho}$ component of the Einstein equations reduces to Eq.~\eqref{no throat condition}, which can be solved as
\begin{align}
 A^{(1)}(\rho) = a_{1} + a_{2} \rho. \label{delta A EFT a1 a2}
\end{align}
Then, one of the throat conditions \eqref{definition of throat} requires $a_{2} = 0$.
By plugging Eq.~\eqref{delta A EFT a1 a2} into the $E^{(1)}{}^{\rho}{}_{\rho} = 0$, we obtain
\begin{align}
 a_{1} = 8 \pi G \mathcal{E},
\end{align}
Thus, the perturbations to the area is a constant determined by the energy--momentum tensor,
\begin{align}
 A^{(1)} = 8 \pi G \mathcal{E}.
\end{align}
Therefore, $F^{(1)}_{\tau \rho}$ is also constant,
\begin{align}
 F^{(1)}_{\tau\rho} = - k q^{(1)}_{e} + k \bar{q}_{e} 8 \pi G \mathcal{E}.
\end{align}
 
Furthermore, ${}^\theta{}_{\theta}$ components of the Einstein equations can be expressed as
\begin{align}
 E^{(1)}{}^{\theta}{}_{\theta} = - \frac{1 + \rho^2}{2 \bar{\ell}^2} \left(
\partial_{\rho}\partial_{\rho} \gamma^{(1)}(\rho) + \frac{4 \rho}{1 + \rho^2} \partial_{\rho} \gamma^{(1)}(\rho) + \frac{2}{1 + \rho^2} \gamma^{(1)}(\rho)
+ 16 \pi G \frac{- 2 \mathcal{E} + \mathcal{P}}{1 + \rho^2}
\right) = 0,
\end{align}
which has the general solution
\begin{align}
 \gamma^{(1)}(\rho) = 8 \pi G (2 \mathcal{E} - \mathcal{P}) + \frac{b_{1}}{1 + \rho^2} + \frac{b_{2} \rho}{1 + \rho^2}.
\end{align}
The matching conditions with the general black hole geometry are automatically satisfied for $c_{4,\gamma} = c_{4,A} = c_{4,F} = 0$.

The combination which appears in the metric components can be evaluated as 
\begin{align}
 \frac{1 + \rho^2}{\gamma(\lambda;\rho)} & = \alpha(\lambda) \left(1 + \beta(\lambda)^2 \left(\rho - \frac{\lambda b_{2}}{2} \right)^2 \right) + \mathcal{O}(\lambda^2),
\end{align}
where we define
\begin{align}
 \alpha(\lambda) &\coloneqq 1 - 8 \pi G \lambda (2 \mathcal{E} - \mathcal{P}) - \lambda b_{1} , \\
 \beta(\lambda)^2 &\coloneqq 1 + \lambda b_{1}.
\end{align}
Then, by introducing new coordinates $\hat{\tau} = \alpha \beta \tau$ and  $\hat{\rho} = \beta (\rho - \lambda b_{2}/2)$, and new scales by 
\begin{align}
 \ell_{1}(\lambda)^2 &= \frac{\ell(\lambda)^2}{\alpha(\lambda) \beta(\lambda)^2} = \ell(\lambda)^2 \left( 1 + 8 \pi G \lambda (2 \mathcal{E} - \mathcal{P})\right) + \mathcal{O}(\lambda^2),\\
 \ell_{2}(\lambda)^2 &= \ell(\lambda)^2 (1 + 8 \pi G \lambda \mathcal{E}), 
\end{align}
 the metric can be expressed as 
\begin{align}
 \bm{g}(\lambda) = \ell_{1}(\lambda)^2 \left( - (1 + \hat{\rho}^2) \bm{d}\hat{\tau}^2 + \frac{\bm{d}\hat{\rho}}{1 + \hat{\rho}^2}\right)
+\ell_{2}(\lambda)^2 \bm{d} \Omega^2 + \mathcal{O}(\lambda^2).
\end{align}
Again, this is the metric for $\text{AdS}^{2} \times \text{S}^2$ and hence possesses the six Killing vectors given by Eqs.~\eqref{xi1}~-~\eqref{J3}, with replacing $\{\tau, \rho\}$ to $\{ \hat{\tau}, \hat{\rho} \}$.
Note that when the matching conditions are satisfied, namely, when $ \mathcal{E} = \mathcal{P}$, we obtain $\ell_{1}^{2} = \ell_{2}^{2}$.

Similarly, the electromagnetic field can be expressed as 
\begin{align}
 \bm{F}(\lambda) &= - k q_{e}(\lambda) (1 - 8 \pi G \lambda \mathcal{E}) \bm{d} \tau \wedge \bm{d} \rho
+ \frac{1}{4 \pi}q_{m}(\lambda) \bm{d} \Omega^2 + \mathcal{O}(\lambda^2) \\
&= - k q_{e}(\lambda) (1 + 8 \pi G \lambda (\mathcal{E} - \mathcal{P}))  \bm{d} \hat{\tau} \wedge \bm{d} \hat{\rho} + \frac{1}{4 \pi} q_{m}(\lambda) \bm{d} \Omega^2 + \mathcal{O}(\lambda^2),
\end{align}
which also has the same symmetry.

To summarize, not only for the background quantities but also the first order quantities possess the symmetry of $\text{AdS}^2 \times \text{S}^2$. Therefore, repeating the same discussion in the previous section, one can show that the energy--momentum tensor in the second-order Einstein equations, which can be covariantly constructed from $\bm{g}$ and $\bm{F}$, possesses the same symmetry even when including the first-order perturbations. As a result, we can conclude that such a energy--momentum tensor does not contribute to the flare-out condition and wormhole solution can not be constructed even for the higher-order perturbations.

\section{No-go theorem for wormholes perturbed from Myers--Perry black holes}
\label{sec:no go MP}
In the following, we apply above discussion for another system: the 5-dimensional Einstein--Hilbert action with a correction term $\mathcal{S}[g]$, that is, the action given by
\begin{align}
 S[\bm{g}] = \int d^5 x  \sqrt{-g} \left( \frac{1}{2 \kappa_{5}^2} R  \right) + \mathcal{S}[\bm{g}]. \label{five dimensional Einstein action}
\end{align}
Here $\kappa_{5}$ is the 5-dimensional gravitational coupling constant, which can be expressed by the 5-dimensional Newton constant $G$ as $ \kappa_{5}^2 = 6 \pi^2 G$.

In the next subsection, we summarize basic properties of the equal-angular-momenta Myers--Perry black hole, a vacuum solution of the 5-dimensional Einstein equations.
Then, in Sec.~\ref{subsec:strategy MP}, we explain the strategy for constructing wormhole solutions as perturbations of the near-horizon geometry of such Myers--Perry black holes. After that, we show no-go theorems for the first-order perturbations in Sec.~\ref{subsec:no go for MP} and for the higher-order perturbations in Sec.~\ref{subsec:no go for MP higher order}. 

\subsection{Myers--Perry black holes and near-horizon, near-extremal geometry}
\label{subsec:MP}
5-dimensional spacetime possesses two rotating axes, and hence the stationary, axisymmetric spacetime is characterized by the time translation symmetry $\mathbb{R}_{t}$ and two rotational symmetries $U(1) \times U(1)$. In the following, we focus on the special case where two independent rotational symmetries $U(1) \times U(1)$ are enhanced into the $U(2) = U(1) \times SU(2)$ symmetry.
In this case, the spacetime is of cohomogeneity-1, and hence the Einstein equations can be reduced to a set of ordinary differential equations. The metric ansatz is given by
\begin{align}
 \bm{g} = - f(r) \mathrm{e}^{2 \Psi(r)} \bm{d}t^2 + \frac{1}{f(r)} 
\bm{d}r^2 + \frac{r^2}{4} \left( \left( \bm{\sigma}_{1}^2 + \bm{\sigma}_{2}^{2} \right)
+ h(r) (\bm{\sigma}_{3} - \Omega(r) \bm{d}t )^2 \right),
\end{align}
where $\bm{\sigma}_{i} ~ (i = 1,2,3)$ are invariant basis for $SU(2)$ symmetries given by
\begin{align}
 \bm{\sigma}_{1} &= - \sin \psi \bm{d} \theta + \sin \theta \cos \psi \bm{d} \phi, \\
 \bm{\sigma}_{2} &= \cos \psi \bm{d} \theta + \sin \theta \sin \psi \bm{d} \phi, \\
 \bm{\sigma}_{3} &= \bm{d} \psi + \cos\theta \bm{d} \phi,
\end{align}
in the coordinates with the Euler angles $\psi \in [0,4\pi], \theta \in [0,\pi], \phi \in [0,2\pi]$. 
The symmetry $\mathbb{R}_{t} \times U(1) \times SU(2)$ can be generated by the following 5 Killing vectors,
\begin{align}
 \bm{\xi}_{t} &= \bm{\partial}_{t}, \\
 \bm{J}_{\psi} &= \bm{\partial}_{\psi} , \\
 \bm{J}_{1} &= \frac{\sin \phi}{\sin \theta} \bm{\partial}_{\psi} + \cos \phi \bm{\partial}_{\theta} - \frac{\sin \phi}{\tan \theta} \bm{\partial}_{\phi}, \\
 \bm{J}_{2} &= \frac{\cos \phi}{\sin \theta} \bm{\partial}_{\psi} - \sin \phi \bm{\partial}_{\theta} - \frac{\cos \phi}{\tan \theta} \bm{\partial}_{\phi}, \\
 \bm{J}_{3} &= \bm{\partial}_{\phi},
\end{align}
which forms the algebras,
\begin{align}
 [\bm{J}_{i}, \bm{J}_{j}] &= \epsilon_{ijk} \bm{J}_{k},
\end{align}
for $i = 1,2,3$, and $\bm{\xi}_{t}, \bm{J}_{\psi}$ commute with every generators. By the definition of the $\text{SU(2)}$ invariant basis, $\bm{\sigma}_{i}$ satisfy
\begin{align}
 \mathsterling_{\bm{J}_{i}} \bm{\sigma}_{j} = 0,
\end{align}
for $i = 1,2,3$. Although $\bm{\sigma}_{1}$ and $\bm{\sigma}_{2}$ themselves are not invariant along $\bm{J}_{\psi}$ direction, the combination
\begin{align}
 \bm{\sigma}_{1}^2 + \bm{\sigma}_{2}^{2} = \bm{d} \theta^2 + \sin^2\theta \bm{d} \phi^2,
\end{align}
is invariant along $\bm{J}_{\psi}$.

The solution to the vacuum Einstein equations with this ansatz is the Myers--Perry solution~\cite{Myers:1986un} with equal-angular-momenta, which is given by
\begin{align}
 f(r) &= 1 - \frac{2 G M}{r^2} + \frac{G^2 a^2}{r^4}, \\
 h(r) &= 1 + \frac{G^2 a^2}{r^4}, \\
 \mathrm{e}^{2 \Psi(r)} &= \frac{1}{1 + \frac{G^2 a^2}{r^4}}, \\
\Omega(r) &= \frac{2 \sqrt{2 G M}}{G a}\frac{\frac{G^2 a^2}{r^4}}{1 + \frac{G^2 a^2}{r^4}}.
\end{align}
Then, the zero points of the function $f(r)$ are obtained as
\begin{align}
 f(r) = 0 \Leftrightarrow r^2 = G M \left(1 \pm \sqrt{1 - \frac{a^2}{M^2}}\right).
\end{align}
We focus on the sub-extremal case $ |a| < M$ and in this case the largest positive root is given by
\begin{align}
 r_{\text{H}} \coloneqq \sqrt{GM\left( 1 + \sqrt{1 - \frac{a^2}{M^2}}\right)}.
\end{align}
Note that $r = r_{\text{H}}$ is the Killing Horizon associated with the Killing vector
\begin{align}
 \bm{\xi}_{\text{H}} = \bm{\xi}_{t} + \Omega_{\text{H}} \bm{\xi}_{\psi},
\end{align}
where $\Omega_{\text{H}}$ is defined by $\Omega_{\text{H}}\coloneqq \Omega(r_{\text{H}})$. This can be confirmed by evaluating the norm of $\bm{\xi}_{\text{H}}$ as
\begin{align}
 g_{\mu\nu} \xi_{\text{H}}^{\mu} \xi_{\text{H}}^{\nu} = - \mathrm{e}^{2 \Psi(r)} f(r) + \frac{r^2}{4} h(r) \left(\Omega(r) - \Omega_{\text{H}} \right)^2,
\end{align}
 which vanishes when $r = r_{\text{H}}$. 

Let us introduce the extremality parameter $\epsilon \in (0,1]$ by
\begin{align}
 \epsilon \coloneqq \sqrt{1 - \frac{a^2}{M^2}}.
\end{align}
$\epsilon \to 0$ corresponds to the extremal black hole and $\epsilon = 1$ corresponds to the Schwarzschild black hole.
We can use the parameters $r_{\text{H}}$ and $\epsilon$ instead of $G M$ and $a/M$, through the relation
\begin{align}
 GM &= \frac{r_{\text{H}}^2}{1 + \epsilon},\\
 \frac{a}{M} &=  \eta \sqrt{1 - \epsilon^2},
\end{align}
where $\eta$ represents the signature of $a$,
\begin{align}
 \eta \coloneqq \text{sgn}(a).
\end{align}
The functions appearing in the metric components are represented as 
\begin{align}
 f(r) &= \frac{1}{r^4} \left( r^2 - r^2_{\text{H}}\right)  \left( r^2 - \frac{1 - \epsilon}{1 + \epsilon} r_{\text{H}}^2\right),\\
 h(r) &= \mathrm{e}^{-2\Psi(r)}=  1 + \frac{1 - \epsilon}{1 + \epsilon} \frac{r_{\text{H}}^4}{r^4},\\
 \Omega(r) &= \Omega_{\text{H}} \frac{2}{1 - \epsilon}\frac{1}{1 + \frac{1+\epsilon}{1 - \epsilon} \frac{r^4}{r_{\text{H}}^4}},
\end{align}
where $\Omega_{\text{H}}$ is given by 
\begin{align}
 \Omega_{\text{H}} \coloneqq \eta \frac{\sqrt{2} \sqrt{1 - \epsilon} }{r_{\text{H}}}.
\end{align}
Note that the value of $h(r)$ at the horizon $r_{\text{H}}$ is given by
\begin{align}
 h_{\text{H}} \coloneqq h(r_{\text{H}}) = \frac{2}{1 + \epsilon}.
\end{align} 

The surface gravity $\kappa$ associated with $\xi_{\text{H}}$, is defined by
\begin{align}
 \nabla_{\rho} (g_{\mu\nu} \xi^{\mu}_{\text{H}} \xi^{\nu}_{\text{H}})|_{r \to r_{\text{H}}} = - 2 \kappa g_{\rho \nu} \xi_{\text{H}}^{\nu}|_{r \to r_{\text{H}}},
\end{align}
and can be evaluated from the relation
\begin{align}
 \kappa = \sqrt{- \frac{1}{2} \nabla_{a} \xi_{\text{H} b} \nabla^{a} \xi_{\text{H}}^{b}}|_{r \to r_{\text{H}}},
\end{align}
as 
\begin{align}
 \kappa = \frac{\sqrt{2}}{\sqrt{1 + \epsilon}} \frac{\epsilon}{r_{\text{H}}} = \sqrt{h_{\text{H}}} \frac{\epsilon}{r_{\text{H}}}.
\end{align}

Then, let us introduce new Rindler-like coordinates,
\begin{align}
 \tau_{\text{R}} &= \kappa t, \\
 \rho_{\text{R}} &= 1 + \frac{2}{\sqrt{h_{\text{H}}} \kappa  r_{\text{H}}} \frac{r^2 - r_{\text{H}}^2}{r_{\text{H}}^2},  \\
 \psi_{\text{R}} &= \psi - \Omega_{\text{H}} t,
\end{align}
or inversely
\begin{align}
 t &= \frac{1}{\kappa} \tau_{\text{R}}, \\
 r &= r_{\text{H}}\sqrt{1 + \frac{\sqrt{h_{\text{H}}} \kappa r_{\text{H}}}{2}  (\rho_{\text{R}} - 1)}, \\
 \psi &= \psi_{\text{R}} + \frac{1}{\kappa} \Omega_{\text{H}} \tau_{\text{R}}.
\end{align}
By introducing the function $\delta$ by
\begin{align}
 \delta \coloneqq \frac{r^2 - r_{\text{H}}^2}{r_{\text{H}}^2} = \frac{r_{\text{H}}\sqrt{h_{\text{H}}}}{2} \kappa (\rho_{\text{R}} - 1) = \frac{\epsilon}{1 + \epsilon} (\rho_{\text{R}} - 1 ) ,
\end{align}
the metric can be expressed as 
\begin{align}
 \bm{g} &= \frac{r_{\text{H}}^2}{4} \Biggl[ - \frac{\rho_{\text{R}}^2 - 1}{1 + (1 + \epsilon) ( \delta  + \frac{1}{2} \delta^2)} \bm{d} \tau_{\text{R}}^2 
+ \frac{1 + \delta}{\rho_{\text{R}}^2 -1} \bm{d} \rho_{\text{R}}^2 + \left(1 + \delta \right)\left( \bm{\sigma}_{\text{R},1}^2 + \bm{\sigma}_{\text{R},2}^2 \right) \notag\\
& \qquad   + \frac{2}{1 + \epsilon} \left( \frac{1 + (1 + \epsilon)(\delta + \frac{1}{2} \delta^2)}{1 + \delta}\right) \left( \bm{\sigma}_{\text{R},3} + \eta \sqrt{1 - \epsilon^2}  \left( \frac{1 + \frac{1}{2} \delta}{1 + (1+\epsilon)(\delta + \frac{1}{2}\delta^2)}\right) (\rho_{\text{R}} - 1) \bm{d}\tau_{\text{R}}  \right)^2 
\Biggr],  \label{exact MP in Rindler}
\end{align}
where we introduce new invariant basis of $SU(2)$ symmetry associated with the new coordinate $\psi_{\text{R}}$ defined by
\begin{align}
 \bm{\sigma}_{\text{R},1} &\coloneqq  - \sin \psi_{\text{R}} \bm{d}\theta + \sin \theta \cos \psi_{\text{R}} \bm{d} \phi,  \\
 \bm{\sigma}_{\text{R},2} &\coloneqq \cos \psi_{\text{R}} \bm{d} \theta + \sin \theta \sin \psi_{\text{R}} \bm{d} \phi, \\
 \bm{\sigma}_{\text{R},3} &\coloneqq \bm{\sigma}_{3} - \Omega_{\text{H}} \bm{d}t = \bm{d} \psi_{\text{R}} + \cos \theta \bm{d} \phi.
\end{align}

The expression \eqref{exact MP in Rindler} is exact and no approximations are used so far. Let us take the near-extremal limit $\epsilon \ll 1$. In this limit, metric can be expressed as
\begin{align}
 \bm{g} &=  \frac{r_{\text{H}}^2}{4} \Biggl[ - \frac{\rho_{\text{R}}^2 - 1}{1 + \delta  + \frac{1}{2} \delta^2} \bm{d} \tau_{\text{R}}^2 
+ \frac{1 + \delta}{\rho_{\text{R}}^2 -1} \bm{d} \rho_{\text{R}}^2 + \left(1 + \delta \right)\left( \bm{\sigma}_{\text{R},1}^2 + \bm{\sigma}_{\text{R},2}^2 \right) \notag\\
& \qquad   + 2 \left( \frac{1 + \delta + \frac{1}{2} \delta^2}{1 + \delta}\right) \left( \bm{\sigma}_{\text{R},3} + \eta \left( \frac{1 + \frac{1}{2} \delta}{1 + \delta + \frac{1}{2}\delta^2}\right) (\rho_{\text{R}} - 1) \bm{d}\tau_{\text{R}}  \right)^2 
\Biggr] + \mathcal{O}(\epsilon).
\end{align}
Then, taking the near-horizon limit $\delta \ll 1$, we obtain
\begin{align}
 \bm{g} &=
\bm{g}^{\text{NHNE}}(1 + \mathcal{O}(\epsilon ,\delta) ), 
\end{align}
with
\begin{align}
 \bm{g}^{\text{NHNE}} \coloneqq \frac{r_{\text{H}}^2}{4} \Biggl[ - (\rho_{\text{R}}^2 - 1) \bm{d} \tau_{\text{R}}^2 
+ \frac{1}{\rho_{\text{R}}^2 -1} \bm{d} \rho_{\text{R}}^2 + \left( \bm{\sigma}_{\text{R},1}^2 + \bm{\sigma}_{\text{R},2}^2 \right)  + 2  \left( \bm{\sigma}_{\text{R},3} + \eta (\rho_{\text{R}} - 1) \bm{d}\tau_{\text{R}}  \right)^2 
\Biggr] . \label{def g NHNE}
\end{align}
One can show that this metric $\bm{g}^{\text{NHNE}}$ is a solution of the vacuum Einstein equations.

For later convenience, let us derive the expression of the near-extremal metric in the $\rho_{\text{R}} \gg 1$ limit, maintaining $\delta \approx \epsilon \rho_{\text{R}} \ll 1$, up to linear order in $\delta \approx \epsilon \rho_{\text{R}}$ and $\epsilon$  terms. In this limit, we obtain
\begin{align}
 \bm{g} &=  \frac{r_{\text{H}}^2}{4} \Biggl[ - \frac{\rho_{\text{R}}^2}{1 + \epsilon \rho_{\text{R}}} \bm{d} \tau_{\text{R}}^2 
+ \frac{1 + \epsilon \rho_{\text{R}}}{\rho_{\text{R}}^2} \bm{d} \rho_{\text{R}}^2 + \left(1 + \epsilon \rho_{\text{R}} \right)\left( \bm{\sigma}_{\text{R},1}^2 + \bm{\sigma}_{\text{R},2}^2 \right) \notag\\
& \qquad \qquad   + 2 \left( \bm{\sigma}_{\text{R},3} + \eta \left( 1 - \frac{1}{2} \epsilon \rho_{\text{R}} \right) \rho_{\text{R}} \bm{d}\tau_{\text{R}}  \right)^2 
\Biggr] + \mathcal{O}\left(\frac{1}{\rho_{\text{R}}}, \epsilon^2, \epsilon (\epsilon \rho_{\text{R}}), (\epsilon \rho_{\text{R}})^2 \right) . \label{g MP NHNE large rho}
\end{align}
We will use this expression for matching the wormhole geometry to black hole geometry.
  
Since the $\tau_{\text{R}}, \rho_{\text{R}}$ part of the metric $\bm{g}^{\text{NHNE}}$, given by Eq.~\eqref{def g NHNE}, is the $\text{AdS}^{2}$ metric in the Rindler-like coordinates, we can express this metric in the global chart through the coordinate transformations \eqref{tau to Rindler} and \eqref{rho to Rindler}, that is, 
\begin{align}
\tau_{\text{R}} &= \tanh^{-1} \left( \sqrt{1 + \frac{1}{\rho^2}} \sin \tau \right), \\
 \rho_{\text{R}} &= \sqrt{1 + \rho^2} \cos \tau. 
\end{align}
In addition, off-diagonal components can be simplified by introducing the new angle $\widehat{\psi}$ given by
\begin{align}
 \psi_{\text{R}} &= \widehat{\psi} + \eta \tanh^{-1}\left( \frac{\rho}{\sqrt{1 + \rho^2}} \frac{1}{\sin \tau}\right) - \eta \tanh^{-1}\left(\frac{\rho}{\tan \tau}\right).
\end{align}
In such new coordinate system $\{\tau, \rho, \theta, \phi, \widehat{\psi}\}$, the metric $\bm{g}^{\text{NHNE}}$ of the near-horizon, near-extremal geometry of the Myers--Perry black hole, Eq.~\eqref{def g NHNE}, can be expressed as 
\begin{align}
 \bm{g}^{\text{NHNE}} = \frac{r_{\text{H}}^2}{4} \left( - (1 + \rho^2 ) \bm{d} \tau^2  + \frac{1}{1 + \rho^2} \bm{d}\rho^2 + \widehat{\bm{\sigma}}_{1}^2 + \widehat{\bm{\sigma}}_{2}^2 + 2 (\widehat{\bm{\sigma}}_{3} + \eta \rho \bm{d} \tau)^2   \right), \label{gMP NH NE}
\end{align}
where $\widehat{\bm{\sigma}}_{i}$ are $SU(2)$ invariant basis associated with the new Euler angles $\{ \theta, \phi, \hat{\psi} \}$. Now, the new coordinate system covers the region beyond the horizon, $\rho \in (- \infty, \infty)$.

Now the symmetry of the equal-angular-momenta Myers--Perry black hole, $\mathbb{R}_{t} \times U(1) \times SU(2)$, is enhanced to $SO(2,1) \times U(1) \times SU(2)$ by the near-horizon, near-extremal limit.
The Killing vectors are explicitly written as follows:

\noindent
$SO(2,1)$:
\begin{align}
  \bm{\xi}_{1} &=  \frac{\rho}{\sqrt{1 + \rho^2}} \cos\tau \bm{\partial}_{\tau} + \sqrt{1 + \rho^2} \sin \tau \bm{\partial}_{\rho} + \eta \frac{1}{\sqrt{1 + \rho^2}} \cos \tau \bm{\partial}_{\widehat{\psi}}, \\ 
  \bm{\xi}_{2} &=  \frac{\rho}{\sqrt{1 + \rho^2}} \sin\tau \bm{\partial}_{\tau} - \sqrt{1 + \rho^2} \cos \tau \bm{\partial}_{\rho} + \eta \frac{1}{\sqrt{1 + \rho^2}} \sin \tau \bm{\partial}_{\widehat{\psi}}, \\
  \bm{\xi}_{3} &=  \bm{\partial}_{\tau}.\\
\end{align}
\noindent
$U(1)$:
\begin{align}
  \widehat{\bm{J}}_{\psi} &= \bm{\partial}_{\widehat{\psi}}.
\end{align}
\noindent
$SU(2)$:
\begin{align}
 \widehat{\bm{J}}_{1} &= \frac{\sin \phi}{\sin \theta} \bm{\partial}_{\widehat{\psi}} + \cos \phi \bm{\partial}_{\theta} - \frac{\sin \phi}{\tan \theta} \bm{\partial}_{\phi}, \\
 \widehat{\bm{J}}_{2} &= \frac{\cos \phi}{\sin \theta} \bm{\partial}_{\widehat{\psi}} - \sin \phi \bm{\partial}_{\theta} - \frac{\cos \phi}{\tan \theta} \bm{\partial}_{\phi}, \\
\widehat{ \bm{J}}_{3} &= \bm{\partial}_{\phi}.
\end{align}
We note that, although the near-horizon, near-extremal geometry corresponds to $\eta = \text{sgn(a)} = \pm 1$ here, the symmetry argument holds for arbitrary constant $\eta$. 

\subsection{Perturbative construction of traversable wormholes }
\label{subsec:strategy MP}
Let us formulate the construction method of traversable wormhole solutions in an analogy of the spherically symmetric case in Sec.~\ref{sec: spherically symmetric wh}.
Thus, let us consider one parameter family of metric $\bm{g}(\lambda)$ and define perturbations by the Taylor expansion with respect to $\lambda$ as 
\begin{align}
 \bm{g}(\lambda) = \bar{\bm{g}} + \lambda \bm{g}^{(1)} + \frac{1}{2} \lambda^2 \bm{g}^{(2)} + \mathcal{O}(\lambda^3),
\end{align}
and solve equations of motion in each order of $\lambda$. We use same notation for any quantities depending on $\lambda$.
As a background metric $\bm{g}$, we consider the metric of the near-horizon, near-extremal geometry of the equal-angular-momenta Myers--Perry black hole given by Eq.~\eqref{gMP NH NE}, renaming the coordinate $\widehat{\psi}$ as $\psi$ and a scale $r_{\text{H}}$ as $\ell$, thus, it can be expressed as 
\begin{align}
 \bar{\bm{g}} = \frac{\ell^2}{4} \left( - (1 + \rho^2 ) \bm{d} \tau^2  + \frac{1}{1 + \rho^2} \bm{d}\rho^2 + \bm{\sigma}_{1}^2 + \bm{\sigma}_{2}^2 + 2 (\bm{\sigma}_{3} + \eta \rho \bm{d} \tau)^2   \right),
\end{align}
and the Killing vectors of $SO(2,1) \times U(1) \times SU(2)$ symmetry for $\bar{\bm{g}}$ are given by $\bm{\xi}_{i}, \bm{J}_{\psi}, \bm{J}_{i}~(i = 1,2,3)$ as given in the previous section, with removing hats. 
We consider stationary perturbations that maintain the angular symmetry, that is, $\mathbb{R}_{\tau} \times U(1) \times SU(2)$ symmetry generated by $\bm{\xi}_{3}, \bm{J}_{\psi}$ and $\bm{J}_{i}$.
Thus, we start with the metric ansatz
\begin{align}
 \bm{g}(\lambda) = \frac{\ell^2}{4} \left( - \frac{(1 + \rho^2 )}{\gamma(\lambda;\rho)} \bm{d} \tau^2  + \frac{\gamma(\lambda;\rho)}{1 + \rho^2} \bm{d}\rho^2 + A(\lambda;\rho) (\bm{\sigma}_{1}^2 + \bm{\sigma}_{2}^2) + 2 B(\lambda;\rho) (\bm{\sigma}_{3} + \Omega(\lambda;\rho) \bm{d} \tau)^2   \right),
\end{align}
with
\begin{align}
\bar{\gamma}(\rho) &= \bar{A}(\rho) = \bar{B}(\rho) = 1, \\
\bar{\Omega}(\rho) &= \eta \rho.
\end{align}
In this setup, the first order metric perturbations can be expressed as 
\begin{align}
 \bm{g}^{(1)} &= \frac{\ell^2}{4} \Biggl(
  \gamma^{(1)}(\rho) (1 + \rho^2)  \bm{d} \tau^2
 + \frac{\gamma^{(1)}(\rho)}{1 +\rho^2} \bm{d}\rho^2  + A^{(1)}(\rho)  (\bm{\sigma}_{1}^2 + \bm{\sigma}_{2}^2)
\notag\\
& \qquad \qquad  + 2 B^{(1)}(\rho) (\bm{\sigma}_{3} + \eta \rho \bm{d} \tau)^2 + 2 \Omega^{(1)}(\rho) \left(( \bm{\sigma}_{3} + \eta \rho \bm{d} \tau) \bm{d}\tau + \bm{d} \tau ( \bm{\sigma}_{3} + \eta \rho \bm{d} \tau) \right) \Biggr).
\end{align}

From the expression of the metric, the area of a $\tau$ and $\rho$-constant surface can be expressed as
\begin{align}
 \mathcal{A}(\lambda;\rho) &= \int_{0}^{\pi} d\theta \int^{2\pi}_{0} d\phi \int_{0}^{4 \pi} d\psi \frac{\ell^3}{8} A(\lambda;\rho) \sqrt{2 B(\lambda;\rho)} \sin\theta = 2\sqrt{2} \pi^2 \ell^3 A(\lambda;\rho) \sqrt{B(\lambda;\rho)} \\
&= 2\sqrt{2} \pi^2 \ell^3 +  2\sqrt{2} \pi^2 \ell^3 \left( A^{(1)} + \frac{1}{2} B^{(1)} \right) \lambda + \mathcal{O}(\lambda^2). \label{area MP}
\end{align}
Then, as in the Reissner--Nordstr\"{o}m case, we demand the following conditions for the perturbations, 
\begin{itemize}
\item \textbf{Throat condition:}  
There is a throat $\rho = \rho_{\text{th}}$ which is defined by
\begin{align}
&\left. \partial_{\rho} \mathcal{A} \right|_{\rho = \rho_{\text{th}}}  = 0, \label{definition of throat MP}
\end{align}
and satisfies flare-out condition
\begin{align}
&\left. \partial_{\rho}^2 \mathcal{A} \right|_{\rho = \rho_{\text{th}}}  > 0. \label{throat condition MP}
\end{align}
If we impose that these conditions are satisfied within the first-order perturbations, using the expression \eqref{area MP}, these conditions can be reduced to 
\begin{align}
  \left. \partial_{\rho} \left(A^{(1)} + \frac{1}{2} B^{(1)}\right) \right|_{\rho = \rho_{\text{th}}} = 0, \label{definition of throat MP first order}
\end{align}
and
\begin{align}
 \left. \partial_{\rho}^2 \left(A^{(1)} + \frac{1}{2} B^{(1)}\right) \right|_{\rho = \rho_{\text{th}}} > 0, \label{throat condition MP first order}
\end{align}
at the first order approximation.
We note that when 
the left hand side of Eq.~\eqref{throat condition MP first order} vanishes, 
the positivity of the higher order contributions in $\partial_{\rho}^2 \mathcal{A}|_{\rho = \rho_{\text{th}}}$ is 
required. In such a case, we can obtain wormhole solution at the higher order in perturbations. 
 \item \textbf{Matching condition:}
We require that the metric $\bm{g}$ in the $\rho \to \infty$ limit, up to leading order correction terms, matches that of the $\rho_{\text{R}} \gg 1, \epsilon \ll 1$ limit, keeping $\delta \approx \epsilon \rho_{\text{R}}\ll 1$, of a near-extremal, near-horizon black hole in the Rindler coordinates up to $\mathcal{O(\delta)}$ terms.
We require similar conditions for $\rho \to - \infty$ limit.
\end{itemize}

In the case where the matched black hole is the exact Myers--Perry black hole, the expression is given by Eq.~\eqref{g MP NHNE large rho}, and the matching conditions for the functions $\gamma, A, B$ and $\Omega$ can be expressed as 
\begin{align}
 \gamma(\lambda; \rho) &\approx  c_{3}(\lambda)^2 (1 + c_{4}(\lambda) \rho ),\label{matching MP gamma} \\
 A(\lambda; \rho) &\approx  c_{3}(\lambda)^2 (1 + c_{4}(\lambda) \rho ),\label{matching MP A} \\
 B(\lambda; \rho) &\approx c_{3}(\lambda)^2,\label{matching MP B} \\
 \Omega(\lambda; \rho) &\approx  c_{3}(\lambda)^{-2} \eta \rho \left(1 - \frac{1}{2} c_{4}(\lambda) \rho\right).\label{matching MP Omega}
\end{align}
in the $\rho \to \infty$ limit. Actually, in this case, one can define the Rindler coordinates $(\tau_{\text{R}}, \rho_{\text{R}})$ by $\tau = c_{1} \tau_{\text{R}}$ and $\rho = c_{2} \rho_{\text{R}}$, with 
\begin{align}
 c_{1} &= \frac{c_{3}^2}{c_{2}}.
\end{align}
In this coordinates, the metric can be expressed as
\begin{align}
 \bm{g} &\stackrel{1 \ll \rho}{\approx}  \frac{c_{3}^2 \ell^2}{4} \Biggl[ - \frac{\rho_{\text{R}}^2}{1 + c_{2} c_{4} \rho_{\text{R}}} \bm{d} \tau_{\text{R}}^2 
+ \frac{1 + c_{2} c_{4} \rho_{\text{R}}}{\rho_{\text{R}}^2} \bm{d} \rho_{\text{R}}^2 + \left(1 + c_{2} c_{4} \rho_{\text{R}} \right)\left( \bm{\sigma}_{1}^2 + \bm{\sigma}_{2}^2 \right) \notag\\
& \qquad \qquad   + 2 \left( \bm{\sigma}_{3} + \eta \left( 1 - \frac{1}{2} c_{2}c_{4} \rho_{\text{R}} \right) \rho_{\text{R}} \bm{d}\tau_{\text{R}}  \right)^2 
\Biggr].
\end{align}
The expression of the metric is equivalent with Eq.~\eqref{g MP NHNE large rho}, with identifying 
\begin{align}
c_{2} c_{4} \equiv \epsilon, \qquad  c_{3} \ell \equiv r_{\text{H}}.
\end{align}

In general, the exterior black hole solution may contain $\mathcal{O}(\lambda)$ corrections from the Myers--Perry black hole. 
To account for this possibility, we relax the matching conditions by allowing each constant to be independent:
\begin{align}
 \gamma(\lambda; \rho) &\approx  c_{3,\gamma}(\lambda)^2 (1 + c_{4,\gamma}(\lambda) \rho ),\label{matching MP gamma gen} \\
 A(\lambda; \rho) &\approx  c_{3,A}(\lambda)^2 (1 + c_{4,A}(\lambda) \rho ),\label{matching MP A gen} \\
 B(\lambda; \rho) &\approx c_{3,B}(\lambda)^2 (1 + c_{4,B}(\lambda) \rho),\label{matching MP B gen} \\
 \Omega(\lambda; \rho) &\approx  \eta c_{3,\Omega}(\lambda)^{-2} \rho \left(1 + c_{4,\Omega}(\lambda) \rho\right),\label{matching MP Omega gen}
\end{align} 
in the $\rho \to \infty$ limit.
The matching to a Myers--Perry black hole, thus, corresponds to $c_{3,\gamma} = c_{3,A} = c_{3,B} = c_{3,\Omega}$ and $c_{4,\gamma} = c_{4,A}= -2 c_{4, \Omega}$, with $c_{4, B} = 0$.
Explicit relations among the overall factors depend on the specific setup.
In the following, we refer to Eqs.~\eqref{matching MP gamma gen} - \eqref{matching MP Omega gen} as the matching conditions.
\subsection{No-go theorem at first order in perturbations}
\label{subsec:no go for MP}
In this section, we show that the no-go theorem for constructing wormhole solutions in the strategy explained in the previous section in the effective field theory approach, where the correction term $\mathcal{S}$ in the action is covariantly constructed from the metric, curvature and their covariant derivatives and are regarded as perturbative amount.
We assign the perturbative expansion parameter $\lambda$ as 
\begin{align}
 \mathcal{S}[\lambda; \bm{g}] = \lambda S^{(1)}[\bm{g}] + \mathcal{O}(\lambda^2).
\end{align}
In this situation, the Einstein equation can be expressed as 
\begin{align}
E^{\mu}{}_{\nu}(\lambda) =  R^{\mu}{}_{\nu}(\bm{g}(\lambda)) - \frac{1}{2} \delta^{\mu}{}_{\nu} R(\bm{g}(\lambda)) - \kappa_{5}^2 \mathcal{T}^{\mu}{}_{\nu}(\lambda; \bm{g}(\lambda)) = 0,
\end{align}
where
\begin{align}
 \mathcal{T}_{\mu\nu}(\lambda; \bm{g}) \coloneqq
 - \frac{2}{\sqrt{-g}}\frac{\delta \mathcal{S}[\lambda; \bm{g}]}{ \delta g^{\mu\nu}},
\end{align}
and can be expanded as 
\begin{align}
 \bar{E}^{\mu}{}_{\nu} &=  \bar{R}^{\mu}{}_{\nu} - \frac{1}{2} \delta^{\mu}{}_{\nu} \bar{R} = 0, \\
 E^{(1)}{}^{\mu}{}_{\nu} &= R^{(1)}{}^{\mu}{}_{\nu} - \frac{1}{2} \delta^{\mu}{}_{\nu} R^{(1)}  - \kappa_{5}^2 \mathcal{T}^{(1)}{}^{\mu}{}_{\nu}(\bar{g}) = 0.
\end{align} 
The important insight here is, again, the symmetry of $\bm{\mathcal{T}}^{(1)}(\bar{g})$: Since it is covariantly constructed from the background metric, curvature and their covariant derivatives, it respects the same symmetry of them.
In our case, $\bar{\bm{g}}$ possesses $SO(2,1) \times U(1) \times SU(2)$ symmetries.
By solving $\mathsterling_{\bm{\xi}_{i}} \bm{\mathcal{T}}^{(1)} = \mathsterling_{\bm{J}_{\psi}} \bm{\mathcal{T}}^{(1)} = \mathsterling_{\bm{J}_{i}} \bm{\mathcal{T}}^{(1)}$ as similar way in the previous section, the general symmetric tensor which respects these symmetries is found to be determined up to three constants $\mathcal{E}, \mathcal{P}_{\theta}, \mathcal{P}_{\psi}$ as follows,
\begin{align}
 \bm{\mathcal{T}}^{(1)} = \mathcal{E} \left( (1 + \rho^2 ) \bm{d} \tau^2  - \frac{1}{1 + \rho^2} \bm{d}\rho^2 \right)
 + \mathcal{P}_{\theta} \left( \bm{\sigma}_{1}^2 + \bm{\sigma}_{2}^2\right)
 + \mathcal{P}_{\psi} (\bm{\sigma}_{3} 
+ \eta \rho \bm{d} \tau)^2.
\end{align}

Then, in the equations of motion $E^{(1)}{}^\tau{}_{\tau} - E^{(1)}{}^{\rho}{}_{\rho} = 0$ and $E^{(1)}{}^{\tau}{}_{\psi} = 0$ the contributions from the energy--momentum tensor are canceled. Hence, we obtain
\begin{align}
0 = E^{(1)}{}^{\tau}{}_{\tau} - E^{(1)}{}^{\rho}{}_{\rho} - \eta \rho E^{(1)}{}^{\tau}{}_{\psi} = \frac{4 ( 1 + \rho^2 )}{\ell^2} \partial_{\rho}^2 \left( A^{(1)} + \frac{1}{2} B^{(1)} \right). \label{tautau - rhorho + taupsi}
\end{align}
Thus, the throat flare-out condition with first-order perturbations, Eq.~\eqref{throat condition MP first order}, can not be satisfied.
This proves the no-go theorem at the first-order perturbations from the equal-angular-momenta Myers--Perry black holes.

\subsection{No-go theorem at higher order in perturbations}
\label{subsec:no go for MP higher order}
Let us proceed the analysis to solve the first order Einstein equations to see whether the throat condition can be satisfied by considering higher-order perturbations.

First, solving Eq.~\eqref{tautau - rhorho + taupsi}, we obtain
\begin{align}
 A^{(1)} + \frac{1}{2} B^{(1)} = a_{1} + a_{2} \rho, \label{A+B}
\end{align}
with integration constants $a_1$ and $a_{2}$.
From one of the throat conditions \eqref{definition of throat MP}, we need to set $a_{2} = 0$.

Then, eliminating $A^{(1)}$ by Eq.~\eqref{A+B}, we obtain
\begin{align}
E^{(1)}{}^{\rho}{}_{\rho} = \frac{4}{\ell^2} \left(\mathcal{E} + B^{(1)} + \eta  \partial_{\rho} \Omega^{(1)} \right) = 0,
\end{align}
thus,
\begin{align}
 \partial_{\rho}  \Omega^{(1)} = - \frac{\mathcal{E} + B^{(1)}}{\eta}.\label{eq Omega dash}
\end{align}
Now the equation of motion $E^{(1)}{}^{\psi}{}_{\phi} = 0$ reduces to the second order differential equation for $B^{(1)}$ given by
\begin{align}
\partial_{\rho}^2 B^{(1)} + \frac{2 \rho}{1 + \rho^2} \partial_{\rho} B^{(1)} - \frac{6}{1 + \rho^2} B^{(1)} = \frac{2 ( 4 \mathcal{E} + 2 \mathcal{P}_{\theta} - \mathcal{P}_{\psi} - 6 a_{1})}{3 (1 + \rho^2)}.
\end{align}
The general solution can be obtained as 
\begin{align}
 B^{(1)} = \frac{6 a_{1} - 4 \mathcal{E} - 2 \mathcal{P}_{\theta} + \mathcal{P}_{\psi}}{9} - a_{3} (1 + 3 \rho^2) +a_{4} \left(3 \rho + (1 + 3 \rho^2) \tan^{-1} \rho \right),
\end{align}
with additional integration constants $a_{3}$ and $a_{4}$.
By using this expression, we can obtain $A^{(1)}$ from Eq.~\eqref{A+B}
and $\Omega^{(1)}$ by integrating Eq.~\eqref{eq Omega dash}, as
\begin{align}
 A^{(1)} &= \frac{12 a_{1} + 4 \mathcal{E} + 2 \mathcal{P}_{\theta} - \mathcal{P}_{\psi}}{18} +\frac{a_{3}}{2}(1 + 3 \rho^2) - \frac{a_{4}}{2}\left(3 \rho + (1 + 3 \rho^2) \tan^{-1} \rho \right),\label{solA general } \\
 \Omega^{(1)} &= a_{5} - \frac{6 a_{1} + 5 \mathcal{E} - 2 \mathcal{P}_{\theta} + \mathcal{P}_{\psi}}{9} \eta \rho + a_{3} \eta \rho ( 1 + \rho^2) - a_{4} \eta \rho \left(\rho + (1 + \rho^2) \tan^{-1} \rho \right), 
\end{align}
where $a_{5}$ is an integration constant.

Now the remaining components of the equations of motion reduces to the single differential equation for $\gamma^{(1)}$ given by
\begin{align}
 &\partial_{\rho}^2 \gamma^{(1)} + \frac{4 \rho}{1 + \rho^2} \partial_{\rho} \gamma^{(1)} + \frac{2}{1 + \rho^2} \gamma^{(1)} \notag\\
&\qquad = \frac{12 a_{1} + 22 \mathcal{E} -16 \mathcal{P}_{\theta}- \mathcal{P}_{\psi}}{9 (1 + \rho^2)}
- 2 \frac{1 + 3 \rho^2}{1 + \rho^2} a_{3}
+ 2 \frac{(3 \rho + (1 + 3 \rho^2)\tan^{-1} \rho)}{1+\rho^2} a_{4}.
\end{align}
The solution can be expressed as 
\begin{align}
 \gamma^{(1)} &= \frac{b_1 + b_{2} \rho}{1 + \rho^2} + \frac{12 a_{1} + 22 \mathcal{E}- 16 \mathcal{P}_{\theta} - \mathcal{P}_{\psi}}{18} \frac{\rho^2}{1 + \rho^2} \notag\\
&\qquad  - \frac{a_{3}}{2} \frac{\rho^2}{1 + \rho^2} ( 2 + \rho^2) + \frac{a_{4}}{2} \left(- \frac{1 - \rho^2}{1 + \rho^2} \rho + (1 + \rho^2) \tan^{-1} \rho \right). 
\end{align}
We have solved all the equations of motion and now the solutions satisfying one of the throat condition \eqref{definition of throat MP} are characterized by the six integration constants $a_{1}, a_{3}, a_{4}, a_{5}, b_{1}$ and $b_{2}$.

Let us now impose the matching conditions~\eqref{matching MP gamma gen}, \eqref{matching MP A gen}, \eqref{matching MP B gen}, and \eqref{matching MP Omega gen}.
In order for the $\rho^2$ terms in Eq.~\eqref{solA general } to be canceled out in the $\rho \to \infty$ limit, we require
\begin{align}
 a_{3} - \frac{\pi}{2} a_{4} = 0.
\end{align}
On the other hand, the same requirement in the opposite limit for $\rho \to - \infty$ limit can be satisfied only if
\begin{align}
 a_{3} + \frac{\pi}{2} a_{4} = 0,
\end{align}
because of the expression $\tan^{-1} \rho \to - \pi/2$ in this limit. Thus, the matching conditions can be simultaneously satisfied only when $a_{3} = a_{4} = 0$.
Note that the Kretschmann invariant is evaluated as 
\begin{align}
 & R_{\mu\nu\rho\sigma} R^{\mu\nu\rho\sigma}  = 
- \frac{96}{\ell^4} \Biggl( 1 + \lambda \biggl( \frac{-12 a_{1} -16 \mathcal{E} + 4 \mathcal{P}_{\theta} + \mathcal{P}_{\psi}}{9} + 2 (1 + 3 \rho^2) a_{3} \notag\\
& \hspace{7cm} - 2 \left(3 \rho + \left(1 + 3 \rho^2\right) \tan^{-1} \rho \right)a_{4} \biggr) + \mathcal{O}(\lambda^2) \Biggr),
\end{align}
indicating that the spacetime cannot be asymptotically regular in both sides unless $a_{3} = a_{4} = 0$.

Under the conditions $a_{3} = a_{4} = 0$, the solutions reduce to
\begin{align}
 \gamma^{(1)} &= \frac{b_1 + b_{2} \rho + b_{3} \rho^2}{1 + \rho^2},\\
 A^{(1)} &= a_{1} - \frac{1}{2} B^{(1)} , \\
 B^{(1)} &= \frac{6 a_{1} - 4 \mathcal{E} - 2 \mathcal{P}_{\theta} + \mathcal{P}_{\psi}}{9},\\
 \Omega^{(1)} &= a_{5} - (\mathcal{E} + B^{(1)}) \eta \rho, 
\end{align}
where we introduce the constant $b_{3}$ defined by
\begin{align}
 b_{3} \coloneqq \frac{12 a_{1} + 22 \mathcal{E}- 16 \mathcal{P}_{\theta} - \mathcal{P}_{\psi}}{18}.
\end{align}
From these expression, one can confirm that all the matching conditions are satisfied with identifying $c_{4,\gamma} = c_{4,A} = c_{4,B} = c_{4,\Omega} = 0$.

Let us write down the metric up to first order in perturbations.
First, one can see that the combination $(1 + \rho^2)/\gamma$ can be expressed as 
\begin{align}
 \frac{1 + \rho^2}{\gamma(\lambda;\rho)} &= (1 + \rho^2) ( 1 - \lambda \gamma^{(1)}) + \mathcal{O}(\lambda^2) \\
&= (1 - \lambda b_{1}) - \lambda b_{2} \rho + (1 - \lambda b_{3}) \rho^2 + \mathcal{O}(\lambda^2) \\
&= \alpha(\lambda) \left(1 + \beta(\lambda)^2 \left(\rho - \frac{1}{2} \lambda b_{2}\right)^2 \right)+ \mathcal{O}(\lambda^2),
\end{align}
where we define
\begin{align}
 \alpha(\lambda) &\coloneqq 1 - \lambda b_{1} + \mathcal{O}(\lambda^2), \\
 \beta(\lambda) &\coloneqq 1 - \frac{\lambda}{2}(b_{3} - b_{1}) + \mathcal{O}(\lambda^2).
\end{align}
Then by introducing new coordinates $\hat{\rho} = \beta (\rho - \lambda b_{2}/2)$ and $\hat{\tau} = \alpha \beta \tau$, we obtain
\begin{align}
 - \frac{1 + \rho^2}{\gamma(\lambda;\rho)} \bm{d}\tau^2 + \frac{\gamma(\lambda;\rho)}{1 + \rho^2} \bm{d} \rho^2 
= \frac{1}{\alpha \beta^2} \left( -  (1 + \hat{\rho}^2) \bm{d}\hat{\tau}^2 + \frac{1}{1 + \hat{\rho}^2} \bm{d} \hat{\rho}^2 \right) + \mathcal{O}(\lambda^2).
\end{align}
Furthermore, other components of the metric can be evaluated as 
\begin{align}
 A(\lambda; \rho) (\bm{\sigma}_{1}^2 + \bm{\sigma}_{2}^2) & = (1 + \lambda A^{(1)}) (\bm{\sigma}_{1}^2 + \bm{\sigma}_{2}^2) + \mathcal{O}(\lambda^2),
\end{align}
and
\begin{align}
 2 B(\lambda;\rho) (\bm{\sigma}_{3} + \Omega(\lambda; \rho) \bm{d} \tau )^2 
&= 2 \left( 1 + \lambda B^{(1)} \right)
\left( \bm{\sigma}_{3} + (\eta \rho - \lambda (\mathcal{E} + B^{(1)}) \eta \rho + \lambda a_{5} ) \bm{d} \tau\right)^2 \notag\\
&= 2 \left( 1 + \lambda B^{(1)} \right)
\left( \widehat{\bm{\sigma}}_{3} + \hat{\eta} \hat{\rho} \bm{d} \hat{\tau}\right)^2.
\end{align}
Here we introduce new angle coordinate
\begin{align}
 \hat{\psi} \coloneqq \psi  + \lambda \left(a_{5} + \frac{\eta b_{2}}{2} \right) \tau,
\end{align}
and
\begin{align}
 \hat{\eta} \coloneqq \eta \frac{1 - \lambda (\mathcal{E} + B^{(1)})}{\alpha \beta^2} = \eta \left( 1 + \lambda \frac{4 \mathcal{E} - 4 \mathcal{P}_{\theta} - \mathcal{P}_{\psi}}{6}\right).
\end{align}

Thus, introducing the scales
\begin{align}
 \ell_{1}^2 &\coloneqq \frac{\ell^2}{\alpha \beta^2} = \ell^2 (1 + \lambda b_{3}) + \mathcal{O}(\lambda^2), \\
 \ell_{2}^2 &\coloneqq \ell^2(1 + \lambda A^{(1)}) + \mathcal{O}(\lambda^2), \\
 \ell_{3}^2 &\coloneqq \ell^2(1 + \lambda B^{(1)}) + \mathcal{O}(\lambda^2),
\end{align}
the metric can be expressed as 
\begin{align}
 \bm{g} = \frac{\ell_{1}^2}{4} \left( - (1 + \hat{\rho}^2) \bm{d} \hat{\tau}^2 + \frac{1}{1 + \hat{\rho}^2} \bm{d} \hat{\rho}^2 \right)
+ \frac{\ell_{2}^2}{4} \left(\bm{\sigma}_{1}^2 + \bm{\sigma}_{2}^2\right)
+ \frac{\ell_{3}^2}{4} 2 \left(\widehat{\bm{\sigma}}_{3} + \hat{\eta} \hat{\rho} \bm{d} \hat{\tau}\right)^2 + \mathcal{O}(\lambda^2).
\end{align}
This metric possesses the enhanced symmetry $SO(2,1) \times U(1) \times SU(2)$ as the background metric. Thus, repeating the discussion similar to the previous subsection, we can conclude that the effective energy--momentum tensor for the second-order perturbations is tightly constrained by this enhanced symmetry and it can not contribute to the throat flare-out condition \eqref{throat condition MP}.
This establishes a no-go theorem for perturbative construction of wormhole geometry from near-horizon geometry of equal-angular-momenta Myers--Perry black holes in the second-order perturbations.
We can repeat similar procedure for arbitrary order of perturbations if we require the matching conditions \eqref{matching MP gamma gen}-\eqref{matching MP Omega gen}.

\section{Summary and Discussion}
\label{sec:summary}
In this paper, we have reformulated the construction of wormhole solutions as perturbations around the near-horizon geometries of black hole spacetimes. 
The perturbative strategy using the Reissner--Nordstr\"{o}m black hole was explained in Sec.~\ref{subsec:strategy RN} and that using the equal-angular-momentum Myers--Perry black hole was discussed  in Sec.~\ref{subsec:strategy MP}.

In the case of the Reissner--Nordstr\"{o}m black hole, we found that when the effective energy--momentum tensor originates from higher-derivative corrections to the Einstein--Maxwell action that are covariantly constructed from the metric, curvature tensor, electromagnetic field strength tensor, and their covariant derivatives, no such perturbative construction works. The key observation is that the effective energy--momentum tensor in this setup is tightly constrained by the enhanced symmetry of near-horizon geometry, namely $SO(2,1) \times SO(3)$ in the Reissner--Nordstr\"{o}m case. Due to this constraint, the form of the effective energy--momentum tensor is determined, up to two constants, as shown in Eq.~\eqref{delta T general}. However, such an energy--momentum tensor does not contribute to satisfying the throat flare-out condition, as demonstrated in Eq.~\eqref{no throat condition}. Furthermore, we confirmed in Sec.~\ref{subsec:no go RN higher} that this structure remains unchanged even at the higher orders in the perturbations. We therefore established a no-go theorem: {\it no wormhole solution can be obtained as a perturbation from the near-horizon geometry of Reissner--Nordstr\"{o}m black holes within the framework of higher-derivative corrections in the effective field theory}.

We also discussed the no-go theorem in the Myers--Perry case in Secs.~\ref{subsec:no go for MP} and \ref{subsec:no go for MP higher order}, where similar symmetry arguments apply. Although the near-horizon geometry possesses a different symmetry structure, $SO(2,1) \times U(1) \times SU(2)$, the essential mechanism leading to the no-go theorem remains the same: the enhanced near-horizon symmetries restrict the form of the effective energy--momentum tensor so severely that the flare-out condition cannot be fulfilled. 
This suggests that the obstruction we identified is not specific to the static, spherically symmetric case, but rather a generic feature of highly symmetric near-horizon geometries.

Let us examine the consistency of our result with known solutions in higher-derivative gravity. 
In Ref.~\cite{Maeda:2008nz}, static, spherically symmetric wormhole solutions in the Gauss--Bonnet gravity with additional matter fields are systematically investigated. There, the solutions were classified into ``GR'' and ``non-GR'' branches, depending on whether the solution remains well-defined in the limit where the Gauss--Bonnet coupling constant goes to zero. It was shown that any solution in the GR branch without cosmological constant requires additional matter that violates the null energy condition.
From our perspective, the solutions in the GR branch can also be interpreted as those within the effective field theory view of higher-derivative corrections.
Therefore, our result, absence of wormhole solutions, is consistent with the conclusion of Ref.~\cite{Maeda:2008nz} in the context of Gauss--Bonnet gravity, though our result can be applied for more general stationary, axisymmetric class of spacetime.

Our no-go theorems clarify the limitations of constructing traversable wormholes within the classical treatment of the perturbative effective field theory framework. It is therefore natural to ask in what directions these limitations might be circumvented.

This may suggest the necessity of explicitly incorporating quantum effects such as the Casimir energy discussed in Ref.~\cite{Maldacena:2018gjk}, which cannot be captured by covariantly constructed higher derivative corrections.

Another possibility is to relax the matching conditions. In particular, in the derivation of the no-go theorem for the Myers--Perry case, the key feature is the vanishing of the constants $a_{3}$ and $a_{4}$, which is required by the matching conditions. A possible extension would be to allow parametrically small but nonzero values of $a_{3}$ and $a_{4}$, so that the terms such as $a_{3} \rho^2$ or $a_{4} \rho^2$ in the $\rho \gg 1$ limit of Eq.~\eqref{solA general } can be consistently matched with the $\mathcal{O}(\delta^2)$ terms in the near-horizon expansion of the black hole geometry.

 Alternatively, one may consider relaxing the strong symmetry of the near-horizon geometries. Our analysis relied crucially on the enhanced symmetry of the near-horizon geometry, which severely constrains the effective energy–momentum tensor. Introducing less symmetric configurations, such as the Kerr black hole in four dimension and the Myers--Perry black holes with unequal angular momenta in five dimensions, may open up new possibilities for satisfying the flare-out condition within EFT framework. We leave these interesting questions for future investigation.

\section*{Acknowledgment}
This work was partially supported by Grants-in-Aid for Scientific Research from the Japan Society for the Promotion of Science (JSPS) under Grant Number 25K07306 (K.M.), and also supported by the Ministry of Education, Culture, Sports, Science and Technology (MEXT) Grant-in-Aid for Transformative Research Areas A Extreme Universe, Grant Numbers~JP21H05186 (K.M.) and JP21H05189 (D.Y.). 

\bibliography{ref}
\bibliographystyle{JHEP.bst}

\end{document}